\documentclass[aoas]{imsart}

\RequirePackage{amsthm,amsmath,amsfonts,amssymb}
\RequirePackage{bm}
\RequirePackage{booktabs}
\RequirePackage[authoryear]{natbib}
\RequirePackage[colorlinks,citecolor=blue,urlcolor=blue]{hyperref}
\RequirePackage{graphicx}
\RequirePackage{pdfpages}

\startlocaldefs
\theoremstyle{plain}

\newtheorem{theorem}{Theorem}
\newtheorem{lemma}[theorem]{Lemma}
\theoremstyle{definition}

\newtheorem{condition}{Condition}

\def\Nu{\mathrm{V}}
\def\pl{\mathrm{pl}}
\DeclareMathOperator*{\var}{var}

\def\bQ{\textbf{Q}}
\def\bX{\textbf{X}}
\def\bU{\textbf{U}}
\def\bO{\textbf{O}}
\def\bbeta{\bm{\beta}}
\def\bgamma{\bm{\gamma}}
\def\btheta{\bm{\theta}}

\allowdisplaybreaks[4]

\endlocaldefs

\begin{document}

\begin{frontmatter}
\title{Semiparametric Analysis of Interval-Censored Data Subject to Inaccurate Diagnoses with A Terminal Event}
\runtitle{Interval-Censored Data Subject to Inaccurate Diagnoses}

\begin{aug}
\author[A]{\fnms{Yuhao}~\snm{Deng}},
\author[A]{\fnms{Donglin}~\snm{Zeng}}
\and
\author[B]{\fnms{Yuanjia}~\snm{Wang}}
\address[A]{Department of Biostatistics,
University of Michigan}
\address[B]{Department of Biostatistics,
Columbia University}
\end{aug}

\begin{abstract}
Interval-censoring frequently occurs in studies of chronic diseases where disease status is inferred from intermittently collected biomarkers. Although many methods have been developed to analyze such data, they typically assume perfect disease diagnosis, which often does not hold in practice due to the inherent imperfect clinical diagnosis of cognitive functions or measurement errors of biomarkers such as cerebrospinal fluid. In this work, we introduce a semiparametric modeling framework using the Cox proportional hazards model to address interval-censored data in the presence of inaccurate disease diagnosis. Our model incorporates sensitivity and specificity of the diagnosis to account for uncertainty in whether the interval truly contains the disease onset. Furthermore, the framework accommodates scenarios involving a terminal event and when diagnosis is accurate, such as through postmortem analysis. We propose a nonparametric maximum likelihood estimation method for inference and develop an efficient EM algorithm to ensure computational feasibility. The regression coefficient estimators are shown to be asymptotically normal, achieving semiparametric efficiency bounds. We further validate our approach through extensive simulation studies and an application assessing Alzheimer's disease (AD) risk. We find that amyloid-beta is significantly associated with AD, but Tau is predictive of both AD and mortality.
\end{abstract}

\begin{keyword}
\kwd{Cox proportional hazards model}
\kwd{Interval censoring}
\kwd{Diagnosis accuracy}
\kwd{Terminal event}
\kwd{Alzheimer's disease}
\end{keyword}

\end{frontmatter}

\section{Introduction}
In studies of chronic diseases, the timing of disease onset or progression is often not directly observed but instead inferred to lie within a time interval between scheduled clinical assessments. This setting gives rise to interval-censored data, which poses unique challenges for survival analysis. Traditional methods for right-censored data are not directly applicable, motivating extensive methodological developments over the past few decades. Early efforts focused on extending the Cox model to accommodate interval censoring using likelihood-based approaches \citep{huang1996efficient, sun1997regression, kooperberg1997hazard, lindsey1998methods, goetghebeur2000semiparametric, kim2003maximum}. More flexible methods, including spline-based estimators, have been proposed to estimate the underlying hazard function \citep{zhang2010spline, wang2016flexible}. More recently, \citet{zeng2016maximum} introduced a semiparametric efficient method for interval-censored data based on the Expectation-Maximization (EM) algorithm, while \citet{gao2019semiparametric} extended the framework to handle multiple types of censored events using nonparametric maximum likelihood approaches.

However, all these methods rely on the assumption of perfect disease diagnosis, that is, they presume that biomarkers or clinical assessments provide accurate and definitive information about whether a disease event has occurred at the time of evaluation. The interval that brackets the disease onset $(t_k,t_{k+1}]$ is implied by the change in the sequences diagnoses $(0,\ldots,0,1,\ldots,1)$ at monitoring times $(t_0,\ldots,t_k,t_{k+1},\ldots,t_N)$. In practice, this assumption often falls short due to imperfect diagnostic tools, measurement error, and the heterogeneity of symptom presentation across individuals. For instance, depending on histopathological severity and clinical criteria, the sensitivity of AD diagnoses ranges from 71\% to 88\%, while specificity varies from 44\% to 71\% \citep{beach2012accuracy, gaugler2013sensitivity}, diabetic events assessed by Hemoglobin A1c have sensitivity and specificity around 70-80\%, while mild cognitive impairment (MCI) evaluated using the Mini-Mental State Examination has reported sensitivity and specificity around 80\% \citep{selvin2011performance, choi2011hemoglobin, mitchell2009meta, arevalo2015mini}. Therefore, the value of diagnosis statuses is not monotonically increasing, and we cannot know the true interval that brackets the disease onset.

In addition to diagnostic inaccuracies, another complication in analyzing interval-censored data arises from the presence of semi-competing risks, where a non-terminal event of interest (e.g., disease onset) may be censored by a terminal event such as death \citep{fine2001semi}. This is particularly relevant in studies of aging and chronic conditions, where mortality can preclude the observation of disease progression. To address the dependencies between non-terminal and terminal events, several joint modeling approaches have been developed for interval-censored settings \citep{leffondre2013interval, jiang2017semi, gao2019semiparametric, ha2020frailty, wei2023bivariate}. However, these models, like earlier methods, typically assume accurate classification of disease events. When diagnostic uncertainty is present, failing to account for this source of variability can mislead inference, leading to inappropriate treatment recommendations and missed opportunities for timely intervention. To account for misdiagnoses in interval-censored data, \citet{pires2021interval} proposed a Bayesian method assuming parametric models, while \citet{yang2024bayesian} further jointly modeled a longitudinal biomarker and considered splines to approximate the baseline hazard function in their Bayesian inference. However, neither of them considered both an interval-censored event and a terminal event, and parametric or spline methods may yield biased estimation if the parametric model is misspecified or spline knots are not chosen appropriately.

This paper introduces a semiparametric modeling framework to analyze interval-censored events in the presence of diagnostic inaccuracies and semi-competing risks. We extend the proportional hazards model to jointly model the hazard functions for the disease event and the terminal event, incorporating shared random effects to capture potential dependence between these two processes. We explicitly adjust for diagnostic inaccuracies by considering sensitivity and specificity in the data likelihood, accounting for uncertainty in the observed disease status, and better reflecting the true timing of the disease event. To estimate model parameters, we adopt a nonparametric maximum likelihood estimation (NPMLE) approach and develop an EM algorithm to handle the complexities introduced by diagnostic misclassification and interval censoring. We establish the asymptotic efficiency of our proposed estimators and demonstrate that they attain the semiparametric efficiency bound. A novel contribution of our work is addressing the mixture likelihood introduced by imperfect diagnosis. Specifically, since the data likelihood derived from observed intervals manifests as a mixture of probabilities, based on where the true event falls within the intervals, this mixture presents substantial theoretical challenges in establishing model identifiability and information operator properties, which are crucial for validating the theoretical integrity of the NPMLE. Additionally, the mixture distribution complicates the NPMLE computation using the EM algorithm, as calculating the conditional expectations of some functions given the observed data need to account for multilevel missingness induced by interval censoring, uncertainty of the intervals containing the event, and latent variables shared between the disease and terminal events.
To demonstrate the practical utility of our proposed framework, we apply it to the Alzheimer’s Disease Neuroimaging Initiative (ADNI) study, a large-scale longitudinal study of aging and cognitive decline. Our analysis identifies several demographic and biomarkers useful for disease risk and survival, such as age, ApoE4 genotype, and levels of amyloid-beta and Tau, and shows the bias introduced by existing approaches that assume perfect diagnosis. 

\section{Motivating Application}
Alzheimer's disease (AD) is a progressive neurodegenerative disorder that primarily affects memory, thinking, and behavior. It is one of the leading causes of death worldwide, and currently has no cure or effective treatment to halt its progression.
The Alzheimer's Disease Neuroimaging Initiative (ADNI) is a comprehensive and widely utilized collection of longitudinal clinical, imaging, genetic, and other biomarker data, first launched in 2004 \citep{petersen2010alzheimer}. The ADNI program recruited individuals aged 55 to 90 years at baseline as an inclusion criterion. A key clinical interest is modeling the risk of AD to facilitate early prevention and intervention. AD diagnosis is based on a combination of neuropsychological tests, brain imaging, and biomarker analysis \citep{chapman2010diagnosis}. Therefore, the occurrence of AD onset is both interval censored and subject to misdiagnosis \citep{crane2012development}. According to existing studies \citep{beach2012accuracy, gaugler2013sensitivity}, the clinical diagnosis has a sensitivity of $p=0.83$ and a specificity of $q=0.55$. 

 We included individuals who had no diagnosis of AD at baseline and had at least one observation of time-varying covariates, resulting in a sample size of $n=759$.  The mean follow-up period was 64.2 months, with a maximum of 233.7 months.
The number of visits ranged from 1 to 19, with an average of 5.8 visits per participant. Among the visits, 11.7\% of the diagnoses were positive. Deaths were recorded in 11.5\% of individuals, of whom 40.2\% underwent autopsy. Postmortem results revealed that 91.4\% of those who received a pathological evaluation had AD before death.



\section{Method} \label{sec:method}

\subsection{Models and data likelihood} \label{sec:model}

We consider data from $n$ independent individuals. For the $i$th subject, we let $T_i$ denote the time to disease and $D_i$ denote the time to the terminal event. Let $\bX_i(t)$ be $ d$-dimensional covariates that are potentially time-varying and assumed to be external. We model the conditional hazard rate of $T_i$ given the covariates and random effects by 
\begin{equation}
\lambda_i(t) = \lambda(t) \exp\{\bbeta^{\top}\bX_i(t)+b_i\},\label{modelT}
\end{equation}
and model the conditional hazard rate function of $D_i$ by 
\begin{equation}
\nu_i(t) = \nu(t) \exp\{\bgamma^{\top}\bX_i(t) + b_i\},\label{modelD}
\end{equation}
where $\lambda(t)$ and $\nu(t)$ are unknown baseline hazard functions, $b_i \sim N(0,\sigma^2)$ is a latent random effect independent of $\bX_i$ with an unknown variance, and we assume $T_i$ and $D_i$ are independent given $\bX_i$ and $b_i$. Therefore, the parameters $\bbeta$ and $\bgamma$ reflect the fixed effects of the covariates, and the random effect $b_i$ captures the dependence between the disease event and terminal event due to unobserved variables. 

We assume that the disease progression is monitored at a sequence of $N_i$ visits, denoted as $\bQ_i = (Q_{i1}, \ldots, Q_{i,N_i})$, and the corresponding diagnosis potentially subject to error are $\bm{\xi}_i = (\xi_{i1}, \ldots, \xi_{i,N_i})$, where $\xi_{ij}=1$ indicates positive diagnosis, meaning that the event is considered as occurring before $Q_{ij}$, otherwise, $\xi_{ij}=0$ for $j=1,\ldots,N_i$.  We further let $Q_{i0} = 0$, $Q_{i,N_i+1} = \infty$ and define intervals $I_{il} = (Q_{il},Q_{i,l+1}]$ for $l=0,\ldots,N_i$. The time to the terminal event, $D_i$, is also subject to right censoring. Therefore, letting $C_i$ be the censoring time, the observed data consist of
$$\{\bO_i = (\bX_i(\cdot),Y_i,\Delta_i,\bQ_i,\bm\xi_i): i=1,\ldots,n\},$$
where $Y_i=\min(D_i, C_i)$ and $\Delta_i=I(D_i\le C_i)$.

We assume that the monitoring times $\bQ_i$ and the censoring time $C_i$ are independent of $T_i$, $D_i$ and $b_i$ given $\bX_i(\cdot)$, and that the diagnosis status $\xi_{ij}$'s are mutually independent given the true disease status for each subject. 
Furthermore, since the diagnosis may not be accurate, we let $p$ and $q$ denote the diagnosis sensitivity and specificity, respectively, and assume them to be known based on the literature. That is, at any time $t$, if $T_i$ occurs before $t$, the probability for a positive diagnosis status at time $t$ is $p$. On the other hand, if $T_i$ occurs after $t$, the probability for a negative diagnosis status at time $t$ is $q$. Given $\bX_i(\cdot)$ and $b_i$, the conditional density of $\bO_i$ is proportional to
\begin{align*}
&\quad~ f(Y_i, \Delta_i\mid \bX_i(\cdot), b_i) \left\{\sum_{l=0}^{N_i} P(T_i\in I_{il}\mid \bX_i(\cdot), b_i) P(\bm\xi_i \mid T_i \in I_{il})\right\}\\
&= f(Y_i, \Delta_i\mid \bX_i(\cdot), b_i) \left\{\sum_{l=0}^{N_i} P(T_i\in I_{il}\mid \bX_i(\cdot), b_i) 
\prod_{j\leq l}q^{1-\xi_{ij}}(1-q)^{\xi_{ij}} \prod_{j>l}p^{\xi_{ij}}(1-p)^{1-\xi_{ij}}\right\},
\end{align*}
where $f(Y_i, \Delta_i\mid \bX_i(\cdot), b_i)$ denotes the conditional density of $(Y_i, \Delta_i)$ given $\bX_i(\cdot)$ and $b_i$.
Under models (\ref{modelT}) and (\ref{modelD}), the likelihood of the observed data is 
\begin{equation}
    \begin{aligned}
     &\prod_{i=1}^{n} \int_{b_i} \nu_i(Y_i)^{\Delta_i} \exp\left\{-\int_0^{Y_i} \nu_i(s)ds\right\}  \sum_{l=0}^{N_i}\bigg[ \prod_{j\leq l}q^{1-\xi_{ij}}(1-q)^{\xi_{ij}} \prod_{j>l}p^{\xi_{ij}}(1-p)^{1-\xi_{ij}} \\ 
    &\qquad\quad \times \left[\exp\left\{-\int_0^{Q_{il}}\lambda_i(s)ds\right\} - I(Q_{i,l+1}<\infty)\exp\left\{-\int_0^{Q_{i,l+1}}\lambda_i(s)ds\right\}\right] \bigg] \\
    &\qquad\quad \times 
    \psi(b_i;\sigma^2)db_i,
    \end{aligned} \label{lik}
\end{equation}
where $\lambda_i(t)$ and $\nu_i(t)$ are given by models (\ref{modelT}) and (\ref{modelD}), respectively, and $\psi(b; \sigma^2)$ denotes the normal density function with mean 0 and variance $\sigma^2$.

\subsection{Nonparametric maximum likelihood estimation and EM algorithm} \label{sec:estimation}

We consider nonparametric maximum likelihood estimation for inference, where we regard the cumulative baseline hazard functions, $\Nu(t)=\int_0^t\nu(s)ds$ and $\Lambda(t)=\int_0^t\lambda(s)ds$, as step functions with jumps at $\mathcal{T}_1 = \{Y_i: i=1,\ldots,n\}$ and  $\mathcal{T}_2 = \{Q_{ij}: i=1,\ldots,n, j=1,\ldots,N_i\}$, respectively. 
More specifically, we sort the elements of $\mathcal{T}_1$ in ascending order as $\{t_{1j}: j=1,\ldots,J_1\}$ and the elements of $\mathcal{T}_2$ in ascending order as $\{t_{2j}: j=1,\ldots,J_2\}$. We use $\nu_{j} = \nu_i\{t_{1j}\}$ and $\lambda_{j} = \lambda\{t_{2j}\}$ to denote the jump size of cumulative baseline hazards at each time point. The nonparametric likelihood is given by
\begin{equation}
    \begin{aligned}
     &\prod_{i=1}^{n} \int_{b_i} \nu\{Y_i\}^{\Delta_i}\exp(\bgamma^{\top}\bX_{i}(Y_i)+b_i)^{\Delta_i} \exp\bigg\{-\sum_{t_{1j} \leq Y_i} \nu_j \exp(\bgamma^{\top}\bX_{i}(t_{1j})+b_i)\bigg\} \\
    &\qquad\quad \times \sum_{l=0}^{N_i} \bigg[\prod_{j\leq l}q^{1-\xi_{ij}}(1-q)^{\xi_{ij}} \prod_{j>l}p^{\xi_{ij}}(1-p)^{1-\xi_{ij}} \\
    &\qquad\qquad\qquad \times \bigg[\exp\bigg\{-\sum_{t_{2j} \leq Q_{il}}\lambda_j\exp(\bbeta^{\top}\bX_{i}(t_{2j})+b_i)\bigg\} \\
    &\qquad\qquad\quad - I(Q_{i,l+1}<\infty)\exp\bigg\{-\sum_{t_{2j} \leq Q_{i,l+1}}\lambda_j\exp(\bbeta^{\top}\bX_{i}(t_{2j})+b_i)\bigg\}\bigg]\bigg] \psi(b_i;\sigma^2) db_i.
    \end{aligned} \label{lik_np}
\end{equation}

To maximize the nonparametric likelihood, we introduce another data generation mechanism that yields the same likelihood function as above. First, following \cite{zeng2016maximum}, we define a sequence of latent Poisson random variables $U_{ij}$ for $i=1,\ldots,n$ and $j=1,\ldots,J_2$ with rate parameter $\lambda_j\exp(\bbeta^{\top}\bX_{i}(t_{2j})+b_i)$. Furthermore, we assume that $\xi_{ij}, j=1,...,N_{ij}$ are mutually independent when conditional on $U_{ij}$'s, and that at each monitoring time $Q_{ij}$, 
\[
P(\xi_{il}=1\mid U_{ij}, j=1,...,J_2) = p I\bigg(\sum_{t_{2j}\le Q_{il}}U_{ij}>0\bigg) + (1-q) I\bigg(\sum_{t_{2j}\le Q_{il}}U_{ij}=0\bigg).
\]
In other words, whenever one of the Poisson variables associated with $t_{2j}\le Q_{il}$ is positive, the positive diagnosis probability is $p$; otherwise, it is $q$. 
The complete data under this new data generation mechanism consist of $\bO_i = (\bX_i(\cdot), Y_i, \Delta_i, \bQ_{i}, \bm\xi_{i})$, $\bU_i = (U_{i1},..., U_{i,J_2})$ and $b_i$ for each subject $i=1,...,n$. 
By simple algebra, we notice 
\begin{align*}
&\quad~ P\bigg(\sum_{t_{2j}\leq Q_{il}}U_{ij}=0, \sum_{t_{2j}\leq Q_{i,l+1}}U_{ij}>0 \mid \bX_i(\cdot), b_i\bigg)\\
&= \exp\bigg\{-\sum_{t_{2j} \leq Q_{il}}\lambda_j\exp(\bbeta^{\top}\bX_{i}(t_{2j})+b_i)\bigg\} \\
&\quad - I(Q_{i,l+1}<\infty)\exp\bigg\{-\sum_{t_{2j} \leq Q_{i,l+1}}\lambda_j\exp(\bbeta^{\top}\bX_{i}(t_{2j})+b_i)\bigg\},
\end{align*}
and based on the assumption for $\xi_{ij}$ under the new data generation mechanism, the joint probability for $(\xi_{ij}, j=1,...,N_i)$ given the event that the first positive Poisson variable in $\bU_i$ occurs in $(Q_{il},Q_{i,l+1}]$, i.e., 
$\sum_{t_{2j}\leq Q_{il}}U_{ij}=0$ and $\sum_{t_{2j}\leq Q_{i,l+1}}U_{ij}>0$, for a specific $l\in\{0,\ldots,N_i\}$ is equal to
\[
P\bigg(\bm\xi_i \bigm| \sum_{t_{2j}\leq Q_{il}}U_{ij}=0, \sum_{t_{2j}\leq Q_{i,l+1}}U_{ij}>0\bigg) = \prod_{j\leq l}q^{1-\xi_{ij}}(1-q)^{\xi_{ij}} \prod_{j>l}p^{\xi_{ij}}(1-p)^{1-\xi_{ij}}.
\]
The information of the latent $\bU_i$ is entirely reflected by $\bm\xi_i$ at monitoring times $\bQ_i$ in the observed data. We conclude that the likelihood function in \eqref{lik_np} is equivalent to the likelihood function for the observed data $\bO_i = (\bX_i(\cdot), Y_i, \Delta_i, \bQ_{i}, \bm\xi_{i})$  under the new data generation mechanism,
\begin{equation}
    \begin{aligned}
     &\prod_{i=1}^{n} \int_{b_i} \nu\{Y_i\}^{\Delta_i}\exp(\bgamma^{\top}\bX_{i}(Y_i)+b_i)^{\Delta_i} \exp\bigg\{-\sum_{t_{1j} \leq Y_i} \nu_j \exp(\bgamma^{\top}\bX_{i}(t_{1j})+b_i)\bigg\} \\
    &\qquad\quad \times \sum_{l=0}^{N_i} \bigg[P\bigg(\bm\xi_i \bigm| \sum_{t_{2j}\leq Q_{il}}U_{ij}=0, \sum_{t_{2j}\leq Q_{i,l+1}}U_{ij}>0\bigg) \\
    &\qquad\qquad\qquad \times P\bigg(\sum_{t_{2j}\leq Q_{il}}U_{ij}=0, \sum_{t_{2j}\leq Q_{i,l+1}}U_{ij}>0 \mid \bX_i(\cdot), b_i\bigg)\bigg] \psi(b_i;\sigma^2) db_i.
    \end{aligned} \label{lik_np_u}
\end{equation}

To maximize the original likelihood function in \eqref{lik_np_u}, we can employ an EM algorithm for the latter by treating $\bU_i$ and $b_i$ as missing data.
 The complete-data likelihood is
\begin{align*}
&\prod_{i=1}^{n} \nu\{Y_i\}^{\Delta_i}\exp(\bgamma^{\top}\bX_i(Y_i)+b_i)^{\Delta_i} \exp\bigg\{-\sum_{t_{1j} \leq Y_i}\nu_j\exp(\bgamma^{\top}\bX_{i}(t_{1j})+b_i)\bigg\} \\
&\qquad \times \prod_{l=1}^{N_i}
\big\{
q^{1-\xi_{ij}}(1-q)^{\xi_{ij}}\big\}^{I\left(\sum_{t_j\le Q_{il}}U_{ij}=0\right)} \big\{p^{\xi_{ij}}(1-p)^{1-\xi_{ij}}\big\}^{I\left(\sum_{t_j\le Q_{il}}U_{ij}>0\right)} \\
&\qquad \times \prod_{j=1}^{J_2} \frac{\{\lambda_j\exp(\bbeta^{\top}\bX_{i}(t_{2j})+b_i)\}^{U_{ij}} \exp\{-\lambda_j\exp(\bbeta^{\top}\bX_{i}(t_{2j})+b_i)\}}{U_{ij}!} \psi(b_i;\sigma^2).
\end{align*}
In the E-step, we need to calculate the conditional expectation of some integrable function of complete data given observed data. We define $S_i$ to be the unique $l\in\{0,\ldots,N_i\}$ such that $\sum_{t_{2j}\leq Q_{il}}U_{ij}=0$ and $\sum_{t_{2j}\leq Q_{i,l+1}}U_{ij}>0$. 
Note that $S_i$ essentially indicates which interval contains $T_i$.  The random variable $S_i$ is a deterministic function of $\bU_i$, which contains sufficient information about $\bU_i$ in the contribution to $\bm\xi_i$. 

Let $R_{il} = Q_{i,l+1}I(Q_{i,l+1}<\infty) + Q_{il}I(Q_{i,l+1}=\infty)$. Later, we will see that estimating the model parameters in the M-step only requires calculating the conditional mean of $U_{ij}I(R_{i,S_i}\geq t_{2j})$ and some functions $g^*(\bO_i,S_i,b_i)$. For the former, simple algebra gives
\begin{align*}
E\{U_{ij}I(R_{i,S_i}\geq t_{2j})\mid \bO_i,S_i,b_i\} &= E\{U_{ij}\mid \bQ_i,\bX_i(\cdot),S_i,b_i\} I(R_{i,S_i}\geq t_{2j}) \\
&= \frac{I(Q_{i,S_i}< t_{2j}\leq R_{i,S_i}) \lambda_j\exp(\bbeta^{\top}\bX_{i}(t_{2j})+b_i)}{1-\exp\left\{-\sum_{Q_{i,S_i}< t_{2r}\leq Q_{i,S_i+1}}\lambda_r\exp(\bbeta^{\top}\bX_{i}(t_{2r})+b_i)\right\}}.
\end{align*}
For 
the posterior mean of any function $g^*(\bO_i,S_i,b_i)$ given observed data $\bO_i$ evaluated at the current parameters, denoted by $\widehat{E}\{g^*(\bO_i,S_i,b_i)\}$, it is calculated by
\[
\widehat{E}\{g^*(\bO_i,S_i,b_i)\} = \frac{\int_{b_i}\sum_{S_i=0}^{N_i}g^*(\bO_i,S_i,b_i){L}(\bO_i,S_i,b_i)db_i}{\int_{b_i}\sum_{S_i=0}^{N_i}{L}(\bO_i,S_i,b_i)db_i},
\]
where
\begin{align*}
L(\bO_i,S_i,b_i) &= \nu\{Y_i\}^{\Delta_i}\exp(\bgamma^{\top}\bX_i(Y_i)+b_i)^{\Delta_i} \exp\bigg\{-\sum_{t_{1j} \leq Y_i}\nu_j\exp(\bgamma^{\top}\bX_{i}(t_{1j})+b_i)\bigg\} \\
&\quad \times \prod_{j\leq S_i}q^{1-\xi_{ij}}(1-q)^{\xi_{ij}} \prod_{j>S_i}p^{\xi_{ij}}(1-p)^{1-\xi_{ij}} \\
&\quad \times \bigg[\exp\bigg\{-\sum_{t_{2j} \leq Q_{i,S_i}}\lambda_j\exp(\bbeta^{\top}\bX_{i}(t_{2j})+b_i)\bigg\} \\
&\qquad - I(Q_{i,S_i+1}<\infty)\exp\bigg\{-\sum_{t_{2j} \leq Q_{i,S_i+1}}\lambda_j\exp(\bbeta^{\top}\bX_{i}(t_{2j})+b_i)\bigg\}\bigg] \psi(b_i;\sigma^2),
\end{align*}
and numerical integration over $b_i$ can be performed based on Gaussian--Hermite quadrature. 

In the M-step, we update $\widehat\bgamma$ by solving
\begin{align*}
&\sum_{i=1}^{n}\Delta_i\left[\bX_i(Y_i)-\frac{\sum_{r=1}^{n}I(Y_r\geq Y_i)\widehat{E}\{\exp(\bgamma^{\top}X_r(Y_i)+b_r)\}X_r(Y_i)}{\sum_{r=1}^{n}I(Y_r\geq Y_i)\widehat{E}\{\exp(\bgamma^{\top}X_r(Y_i)+b_r)\}}\right] = 0
\end{align*}
and update $\widehat\bbeta$ by solving
\begin{align*}
&\sum_{i=1}^{n}\sum_{j=1}^{J_2}\widehat{E}\{U_{ij}I(R_{i,S_i}\geq t_{2j})\} \\
&\qquad \times\bigg[\bX_{i}(t_{2j})-\frac{\sum_{r=1}^{n}\widehat{E}\{I(R_{r,S_i}\geq t_{2j})\exp(\bbeta^{\top}X_{r}(t_{2j})+b_r)\}X_{r}(t_{2j})}{\sum_{r=1}^{n}\widehat{E}\{I(R_{r,S_i}\geq t_{2j})\exp(\bbeta^{\top}X_{r}(t_{2j})+b_r)\}}\bigg] = 0
\end{align*}
using one-step Newton--Raphson. The variance of the random effect is updated by
$
\widehat\sigma^2 = n^{-1}\sum_{i=1}^{n}\widehat{E}(b_i^2).
$
Finally, we update the jump size of the cumulative baseline hazards by
\begin{align*}
\widehat\nu_j &= \frac{\sum_{i=1}^{n}\Delta_iI(Y_i=t_{1j})}{\sum_{i=1}^{n}I(Y_i\geq t_{1j})\widehat{E}\{\exp(\widehat\bgamma^{\top}\bX_{i}(t_{1j})+b_i)\}}, \\
\widehat\lambda_j &= \frac{\sum_{i=1}^{n}\widehat{E}\{U_{ij}I(R_{i,S_i}\geq t_{2j})\}}{\sum_{i=1}^{n}\widehat{E}\{I(R_{i,S_i}\geq t_{2j})\exp(\widehat\bbeta^{\top}\bX_{i}(t_{2j})+b_i)\}}.
\end{align*}
We repeat the E-step and M-step until convergence. 

One significant advantage of using the EM algorithm is that the update for high-dimensional nuisance parameters, both $\lambda$ and $\nu$, has an explicit form in each iteration, and the regression coefficients are updated via the Newton--Raphson algorithm. Therefore, this computing algorithm is more reliable than any black-box optimization. 
We let $\widehat\btheta = (\widehat\bbeta^{\top},\widehat\bgamma^{\top},\widehat\sigma^2)^{\top}$ and $\widehat{\mathcal{A}} = (\widehat\Nu(\cdot),\widehat\Lambda(\cdot))$ be the estimates of $\btheta = (\bbeta^{\top},\bgamma^{\top},\sigma^2)^{\top}$ and $\mathcal{A} = (\Nu(\cdot),\Lambda(\cdot))$, respectively.

\subsection{Extension to incorporate accurate pathological diagnoses} \label{sec:extension}

An accurate pathological diagnosis for AD can be obtained in certain individuals by autopsy at the time of death. Incorporating postmortem pathological diagnosis can therefore enhance the precision of model parameter estimates. Here, we assume that autopsies are performed on a random sample of individuals after death. We define $G_i$ as an indicator of whether an autopsy is performed on the individual. For a subject with an autopsy $(G_i=1)$, we introduce $W_i$ to represent the autopsy result: $W_i = 1$ indicates that the disease was present before $Y_i$, whereas $W_i = 0$ otherwise. Consequently, the observed data now consist of $\{\bO_i = (\bX_i(\cdot),\bQ_i,\bm\xi_i,Y_i,\Delta_i,\Delta_iG_i, \Delta_iG_iW_i): i=1,\ldots,n\}$.

Pathological diagnosis reveals the true status of the disease at $Y_i$, which functions as a gold standard assessment at $Y_i$ with perfect sensitivity and specificity of 100\%. If an autopsy reveals a positive disease status, $T_i$ can belong to any interval $I_{il}$ where $l=0,\ldots,N_i$, but it must be less than $Y_i$. In contrast, if the autopsy indicates a negative disease status, $T_i$ must be greater than $Y_i$. We define $Q_{il}^* = Q_{il}$ for $l=0,\ldots,N_i$, and $Q_{i,N_i+1}^* = \infty(1-\Delta_iW_i) + Y_i\Delta_iG_i$. Based on the arguments above, the likelihood of observed data is given by
\begin{equation}
    \begin{aligned}
    &\prod_{i=1}^{n} \int_{b_i} \nu_i(Y_i)^{\Delta_i} \exp\bigg\{\int_0^{Y_i} \nu_i(s)ds\bigg\} \\
    &\qquad \times \sum_{l=0}^{N_i} \bigg[\prod_{j\leq l}q^{1-\xi_{ij}}(1-q)^{\xi_{ij}} \prod_{j>l}p^{\xi_{ij}}(1-p)^{1-\xi_{ij}} \bigg[\exp\bigg\{-\int_0^{Q_{il}^*}\lambda_i(s)ds\bigg\} \\
    &\qquad\qquad - I(Q_{i,l+1}^*<\infty) \exp\left\{-\int_0^{Q_{i,l+1}^*}\lambda_i(s)ds\right\}\bigg]\bigg]^{1-\Delta_iG_i(1-W_i)} \\
    &\qquad \times \bigg[\prod_{j\leq N_i}q^{1-\xi_{ij}}(1-q)^{1-\xi_{ij}} \exp\bigg\{-\int_0^{Y_i}\lambda_i(s)ds\bigg\}\bigg]^{\Delta_iG_i(1-W_i)}
    \psi(b_i;\sigma^2)db_i.
    \end{aligned}
\end{equation}

We can apply the same EM algorithm to calculate the nonparametric maximum likelihood estimators, except that for subject $i$ who received an autopsy, we expand their interval to include $(Q_{i,N_i}, Y_i]$ and $(Y_i, \infty)$ to incorporate the additional diagnosis at $Y_i$. Furthermore, the sensitivity and specificity at $Y_i$ are both equal to 100\%.

\subsection{Model prediction} \label{sec:prediction}

To predict the cumulative incidence function of the disease and the survival probability for the whole population,  we can use the following plug-in estimators
\begin{align*}
\widehat{F}(t) &= \frac{1}{n}\sum_{i=1}^{n} \int_b \int_0^t \exp\left\{-\int_0^s\exp(\widehat\bbeta^{\top}\bX_i(u)+b)d\widehat\Lambda(u) -\int_0^s\exp(\widehat\bgamma^{\top}\bX_i(u)+b)d\widehat\Nu(u)\right\} \\
&\qquad\qquad\qquad \times \exp(\widehat\bbeta^{\top}\bX_i(s)+b) \psi(b;\widehat\sigma^2)d\widehat\Lambda(s)db, \\
\widehat{S}(t) &= \frac{1}{n}\sum_{i=1}^{n} \int_b \exp\left\{-\int_0^t\exp(\widehat\bgamma^{\top}\bX_i(s)+b)d\widehat\Nu(s)\right\}\psi(b;\widehat\sigma^2)db.
\end{align*}
Furthermore, the proposed modeling approach with random effects is particularly useful for predicting the disease and death risks based on historical information for the same subject. For example, at a given time $t$, given the historical diagnoses data $\bQ=(Q_1,\ldots,Q_k)$ and $\bm\xi=(\xi_1,\ldots,\xi_k)$ with $Q_k \leq t$, we can predict the subject's future survival probabilities and cumulative incidence of the disease, assuming that the covariate value is $\bX$. Specifically, we define $Q_0=0$, $Q_{k+1}=\infty$, and
\begin{align*}
\widehat{P}(D>t \mid \bX,b) &= \exp\left\{-\int_0^t\exp(\widehat\bgamma^{\top}\bX(s)+b)d\widehat\Nu(s)\right\}, \\
\widehat{P}(\bm\xi\mid Q_l< T\leq Q_{l+1}) &= \prod_{j\leq l}q^{1-\xi_j}(1-q)^{\xi_j}\prod_{j>l}p^{\xi_j}(1-p)^{1-\xi_j}, \\
\widehat{P}(Q_l< T\leq Q_{l+1}\mid \bX,b) &= \exp\left\{-\int_0^{Q_{l}}\exp(\widehat\bbeta^{\top}\bX(s)+b)d\widehat\Lambda(s)\right\} \\
&\quad - I(Q_{l+1}<\infty)\exp\left\{-\int_0^{Q_{l+1}}\exp(\widehat\bbeta^{\top}\bX(s)+b)d\widehat\Lambda(s)\right\}. 
\end{align*}
Then the survival probability at time $t^* \geq t$ is estimated by
\begin{align*}
&\quad ~ \widehat{P}(D>t^* \mid \bQ, \bm\xi, \bX, D>t) \\
&= \frac{\int_b \widehat{P}(D>t^*\mid \bX,b)\sum_{l=0}^{k}\widehat{P}(\bm\xi \mid Q_l< T\leq Q_{l+1})\widehat{P}(Q_{l}< T\leq Q_{l+1} \mid \bX,b)\psi(b;\widehat\sigma^2)db}{\int_b \widehat{P}(D>t\mid \bX,b)\sum_{l=0}^{k}\widehat{P}(\bm\xi \mid Q_l< T\leq Q_{l+1})\widehat{P}(Q_l< T\leq Q_{l+1} \mid \bX,b)\psi(b;\widehat\sigma^2)db},
\end{align*}
and the disease-free survival probability (the probability that an individual survives and does not have disease) at $t^* \geq t$ is estimated by
\begin{align*}
&\quad ~ \widehat{P}(D>t^*, T>t^* \mid \bQ, \bm\xi, \bX, D>t) \\
&= \frac{\int_b \widehat{P}(D>t^*\mid \bX,b)\widehat{P}(\bm\xi \mid Q_k< T\leq Q_{k+1})\widehat{P}(T>t^* \mid \bX,b)\psi(b;\widehat\sigma^2)db}{\int_b \widehat{P}(D>t\mid \bX,b)\sum_{l=0}^{k}\widehat{P}(\bm\xi \mid Q_l< T\leq Q_{l+1})\widehat{P}(Q_l< T\leq Q_{l+1} \mid \bX,b)\psi(b;\widehat\sigma^2)db}.
\end{align*}

\section{Asymptotic Theory} \label{sec:asymptotics}

Let $\btheta_0 = (\bbeta_0^{\top},\bgamma_0^{\top},\sigma_0^2)^{\top}$ be the true value of the parametric part and $\mathcal{A}_0 = (\Nu_0(\cdot),\Lambda_0(\cdot))$ be the true baseline cumulative hazards. Let $\tau$ be the end of the study. 
We establish the asymptotic properties of $(\widehat\btheta,\widehat{\mathcal{A}})$ under the following regularity conditions.

\begin{condition}
The true parameter $\btheta_0$ is an interior point of a known compact set $\Theta$. The cumulative baseline hazards $\Lambda_0(t)$ and $\Nu_0(t)$ are strictly increasing and continuously differentiable on $[0,\tau]$ with $\Lambda_0(0)=\Nu_0(0)=0$, and $\Lambda_0(\tau)<M$ for a known constant $M$.
\end{condition}
\begin{condition}
With probability 1, $\bX_i(t)$ has bounded total variation on $[0,\tau]$. If there exists a constant vector $\bm\alpha$ and a deterministic function $c(t)$ such that  $\bm\alpha^{\top}\bX_i(t)=c(t)$ for any $t\in[0,\tau]$ with probability 1, then $\bm\alpha=0$ and $c(t)=0$ for any $t\in[0,\tau]$.
\end{condition}
\begin{condition}
There exists a constant $\delta>0$ such that $P(Y_i=\tau \mid \bX_i) > \delta$ almost surely.
\end{condition}
\begin{condition}
The monitoring times $\bQ_i$ have finite support $\mathcal{Q}$ with the least upper bound $\tau$. The number of potential monitoring times $N_i$ is positive with $E(N_i)<\infty$. There exists a positive constant $\eta$ such that $Q_{i,l+1}-Q_{il} \geq \eta$ for every $l \in \{0,\ldots,N_i-1\}$ almost surely.
The Radon--Nikodym derivative of $(Q_{i,l+1},Q_{il})$ with respect to some measure $\mu\time\mu$ on $[0,\tau]\times[0,\tau]$ conditional on $N_i$ and $\bX_i(\cdot)$ exists, denoted by $f_{l,l+1}(u,v;N_i,\bX_i(\cdot))$, is positive and twice-continuously differentiable.
\end{condition}
\begin{condition}
The sensitivity and specificity $p+q \neq 1$.
\end{condition}

Conditions 1 and 2 are standard conditions for failure time regression with time-dependent covariates \citep{zeng2016maximum, gao2019semiparametric}. The time-varying covariates should not be linearly dependent for the identification of parameters. In Condition 1, assuming a known upper bound for $\Lambda_0(\tau)$ is necessary to ensure that its estimator does not diverge in the consistency proof; however, we will show in the proof that this assumption is not needed when postmortem data are available. Condition 3 states that there is a positive probability that there are individuals alive at the end of the study, likely due to administrative censoring in practice. Condition 4 states that two adjacent monitoring times are separated by at least $\eta$, otherwise the data may contain exact observations. If the monitoring times are continuous random variables on $[0,\tau]$, the dominating measure $\mu$ can be chosen as the Lebesgue measure. The number of monitoring times can either be fixed or random across individuals. Condition 5 ensures that the diagnoses contain information about disease onset, so the posterior probability that the disease occurs in each interval differs. If $p+q=1$, switching positive and negative diagnosis results will lead to the same likelihood, so the model parameters cannot be identified. We note that for most disease biomarkers, $p$ and $q$ are both larger than 0.5.

\begin{lemma} \label{lem1}
Under Conditions 1--5, if the density function of the observed data $p(\btheta_*,\mathcal{A}_*) = p(\btheta_0,\mathcal{A}_0)$ for some parameters $(\btheta_*,\mathcal{A}_*)$ with probability 1, then $\btheta_*=\btheta_0$ and $\mathcal{A}_*=\mathcal{A}_0$ in $[0,\tau]\times\mathcal{Q}$.
\end{lemma}

Lemma \ref{lem1} ensures that the solution is unique. The strong consistency of $(\widehat\btheta,\widehat{\mathcal{A}})$ and the weak convergence of $\widehat\btheta$ are further shown in the following theorems.

\begin{theorem} \label{thm1}
Let $\|\cdot\|$ be the Euclidean norm and  $\|\cdot\|_{\infty}$ be the sup-norm on $[0,\tau]\times \mathcal{Q}$. Under Conditions 1--5, $\|\widehat\btheta-\btheta_0\| + \|\widehat{\mathcal{A}}-\mathcal{A}_0\|_{\infty} \to_p 0$.
\end{theorem}

\begin{theorem} \label{thm2}
Under Conditions 1--5, $\sqrt{n}(\widehat\btheta-\btheta_0)$ converges to a mean-zero normal distribution, whose variance attains the semiparametric efficiency bound.
\end{theorem}

The proofs of these theorems are provided in Supplementary Materials A, B, and C. For interval-censored data, similar to \citep{huang1996efficient}, we show that the convergence rate of the estimated baseline cumulative hazards $\widehat{\mathcal{A}}$ is $O_p(n^{-1/3})$, while the convergence rate of the estimated parametric compoent $\widehat\btheta$ is $O_p(n^{-1/2})$ in our method. Nevertheless, we can make statistical inference on the parameters $\widehat\btheta$. The asymptotic variance of $\widehat\btheta$ can be estimated by profile likelihoods \citep{murphy2000profile}. More specifically, let $L_i(\bO_i;\btheta,\mathcal{A})$ be the likelihood contributed by the $i$th individual at the parameter value $(\btheta,\mathcal{A})$. For any fixed $\btheta$, we let $\widehat {\mathcal{A}}_{\btheta}$ be the NPMLE for $\mathcal{A}$ using the same EM algorithm as before, except that $\btheta$ is fixed in the iterations. We define 
$
\pl_i(\btheta) =  \log L_i(\bO_i;\btheta,\widehat{\mathcal{A}}_{\btheta}),
$
and let $\bm{e}_j$ be a $d$-dimensional unit vector with only the $j$th element equal to 1. Then the score function for subject $i$ is numerically evaluated by
\[
\mathcal{S}_i(\widehat\btheta) = \frac{1}{h_n} \left(\begin{array}{c} 
\pl_i(\widehat\btheta + h_n \bm{e}_1) - \pl_i(\widehat\btheta) \\ \vdots \\ \pl_i(\widehat\btheta + h_n \bm{e}_d) - \pl_i(\widehat\btheta)
\end{array}\right),
\]
where $h_n$ is a constant of order $O(1/\sqrt{n})$. The asymptotic variance of $\widehat\btheta$ is numerically estimated by
$
\widehat{\var}(\widehat\btheta) = \left\{\sum_{i=1}^{n} \mathcal{S}_i(\widehat\btheta)\mathcal{S}_i(\widehat\btheta)^{\top}\right\}^{-1}.
$

\section{Simulation Studies} \label{sec:simulation}

We considered $n = 500$ in our simulation studies. A baseline covariate was generated from $X_{1i} \sim U(-1,1)$ and a time-varying covariate was generated from $X_{2i}(t) = c_i \sin(0.2a_it+u_i)$, where $a_i,u_i,c_i \sim U(-1,1)$ were independent. The time-varying covariate mimicked some fluctuating biomarker measurements over time. The variance of the random effect $\sigma^2 = 0.25$. The events were generated from
$\Lambda_i(t) = \log(1+0.25t)e^{0.5X_{1i}+0.5X_{2i}(t)+b_i}$, and $\Nu_i(t) = 0.01t^2e^{0.5X_{1i}-0.5X_{2i}(t)+b_i}$.
We assumed that each individual had six scheduled visits, with scheduled time $Q_{ij} = Q_{i,j-1}+0.1+U_{ij}$ ($j=1,\ldots,6$), where $Q_{i0}=0$ and $U_{ij} \sim U(0,2)$. The diagnosis $\xi_{ij}$, provided that the $i$th individual is still alive at $Q_{ij}$, was subject to measurement error, with sensitivity $p \in \{0.9,0.8\}$ and specificity $q \in \{0.6,0.5\}$. The censoring time $C_i$ was the minimum of $U(3,9)$ and $6$. The proportion of observing the death event is about 32\%, and the proportion of having a disease before death or censoring is about 56\%. If death occurs before censoring, we assumed the probability to receive an autopsy, i.e., $r = P(G_i=1 \mid \Delta_i=1)$, could vary from 0, 0.5 to 1, corresponding to none, half, and all of the death subjects having a postmortem diagnosis.

Computationally, the initial value of parameters was set at 0 for elements of $\bbeta$ and $\bgamma$ and $0.2$ for $\sigma^2$. The jump size of baseline hazards was set at $1/|\mathcal{T}_1|$ for $\Nu(\cdot)$ and $1/|\mathcal{T}_2|$ for $\Lambda(\cdot)$. Convergence of the EM algorithm was claimed if the absolute difference of estimates in two adjacent iterations (tolerance) was smaller than $5\times10^{-4}$, and on average, it usually took less than 200 iterations to converge in all settings. When calculating the standard error, we set $h_n=n^{-1/2}$. We also tried $h_n=5n^{-1/2}$, and the numerical results were similar.
Since existing methods for analyzing such data all assumed the perfect diagnosis, we considered the following three alternatives when comparing with our approach: (1) using the visit of the first positive diagnosis and the previous visit as the true interval (First-Diag), (2) using the last diagnosis or pathological evaluation to obtain current status data for the disease event (Last-Diag), and (3) assuming no random effect by setting $\sigma^2=0$ in our method (NoRandEff). For the first two alternatives, we applied the proposed method in \citet{gao2019semiparametric} for parameter estimation. 

In Table \ref{tab:simu1}, we show the average bias, standard deviation (SD), average standard error (SE) of the estimates $\widehat\btheta$, and coverage percentage (CP) of the nominal 95\% confidence intervals based on 1000 replicates. 
We find that the average bias of our estimates is negligible, and the coverage percentage is close to 95\% in most settings. The standard deviation is smaller with a higher sensitivity $p$, specificity $q$, and probability of receiving an autopsy $r$. Higher sensitivity and specificity imply that the diagnosis is more accurate, and a higher probability of receiving an autopsy provides more information on the true disease status, so there is less uncertainty in inferring the interval that covers the disease onset. In contrast, the estimates by competing methods (First-Diag and Last-Diag) are severely biased since they use incorrect diagnosis information. The method ignoring random effects (NoRandEff) shows larger bias, especially in estimating $\bgamma$, and the coverage percentage is generally lower than the nominal level.

\begin{figure}[!tbh]
\centering
\includegraphics[width=0.8\textwidth]{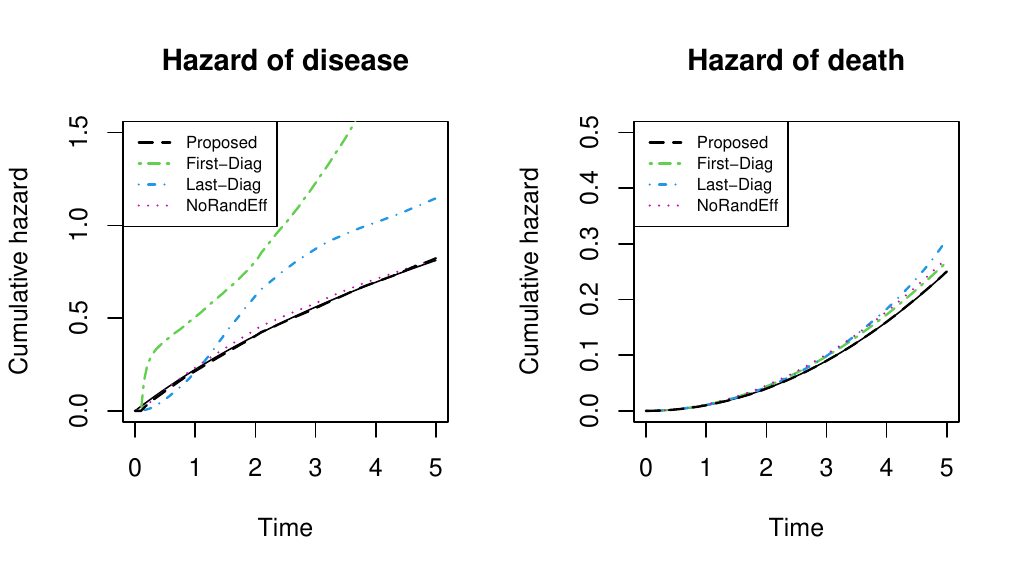}
\caption{The average estimated cumulative baseline hazards with $p=0.9$, $q=0.6$, $r=0.5$ and $n=500$. The true values are plotted in solid lines. The average estimated curves are plotted in dotted/dashed lines.} \label{fig:simu_haz}
\end{figure}

In Setting 2 ($p=0.9$, $q=0.6$, $r=0.5$), the median number of iterations to achieve convergence is 126, and the median computation time is 26 minutes for the proposed method in R (version 4.4.0) on an Intel CPU  with 3GHz.
In Figure \ref{fig:simu_haz}, we show the average estimated cumulative baseline hazards $\widehat\Lambda(t)$ and $\widehat\Nu(t)$ in dashed lines with true values in solid black lines. Our method shows negligible bias, while the competing methods show considerable bias. In our data generation, the specificity of diagnosis is much lower than the sensitivity, so a healthy individual is likely to be classified as diseased, and the disease risk is overestimated. The First-Diag method tends to classify individuals as diseased at early follow-ups, resulting in a very high estimated disease hazard. In contrast, the Last-Diag method exhibits bias at later follow-ups. Since the competing methods yield biased estimates for the magnitude of random effects, the estimated death hazard is also biased. 

In Supplementary Material E, we present the simulation results for a larger sample size $n=1,000$ in Table S1. The standard deviations of the estimated parameters decrease with an increase in sample size. The standard deviation and standard error exhibit better agreement under this scenario, given a larger sample size. We also conduct two additional simulation studies to assess the influence of the convergence criterion and initial values, as shown in Tables S2 and S3. When using a stricter convergence criterion, with tolerance $<1\times10^{-5}$, the median number of iterations required to achieve convergence is approximately 4.3 times that of tolerance $<5\times10^{-5}$. When randomly choosing the initial values from a distribution, such as $\beta_1 \sim N(0.5,1)$, $\beta_2 \sim N(0.5,1)$, $\gamma_1 \sim N(0.5,1)$, $\gamma_2 \sim N(-0.5,1)$, and $\sigma^2 \sim Gamma(2,4)$, we find that the estimation results are virtually the same. Hence, they are insensitive to the initial values.

\begin{table}
\centering
\caption{Summary of the simulation results for $n=500$} \label{tab:simu1}
\setlength{\tabcolsep}{2.5pt}
\begin{tabular}{cccccccccccccccccccc}
\toprule
& \multicolumn{4}{c}{Proposed method} & & \multicolumn{4}{c}{First-Diag} & & \multicolumn{4}{c}{Last-Diag} & & \multicolumn{4}{c}{NoRandEff}  \\
 \cmidrule(lr){2-5} \cmidrule(lr){7-10} \cmidrule(lr){12-15} \cmidrule(lr){17-20}
 & Bias & SD & SE & CP &  & Bias & SD & SE & CP &  & Bias & SD & SE & CP &  & Bias & SD & SE & CP \\ 
 \midrule
  \multicolumn{20}{c}{\underline{Setting 1: $p=0.9$, $q=0.6$, $r=1$}} \\
  $\beta_1$ & .011 & .152 & .151 & .946 &  & -.372 & .097 & .098 & .033 &  & -.237 & .111 & .112 & .418 &  & -.030 & .138 & .131 & .934 \\ 
  $\beta_2$ & .023 & .296 & .283 & .939 &  & -.238 & .184 & .180 & .726 &  & -.171 & .205 & .208 & .863 &  & -.027 & .268 & .207 & .864 \\ 
  $\gamma_1$ & .002 & .160 & .157 & .946 &  & -.017 & .153 & .154 & .951 &  & -.022 & .158 & .155 & .947 &  & -.021 & .152 & .140 & .930 \\ 
  $\gamma_2$ & .003 & .264 & .269 & .957 &  & .023 & .253 & .264 & .961 &  & .028 & .262 & .265 & .959 &  & .026 & .251 & .205 & .882 \\ 
  $\sigma^2$ & .013 & .132 & .138 & .964 &  & -.207 & .013 & .102 & .380 &  & -.131 & .039 & .116 & 1.000 &  & --& -- & -- & -- \\
  \multicolumn{20}{c}{\underline{Setting 2: $p=0.9$, $q=0.6$, $r=0.5$}} \\
  $\beta_1$ & .014 & .161 & .163 & .949 &  & -.357 & .099 & .098 & .056 &  & -.248 & .119 & .122 & .460 &  & -.032 & .146 & .140 & .929 \\ 
  $\beta_2$ & .023 & .316 & .303 & .946 &  & -.320 & .187 & .181 & .561 &  & -.205 & .224 & .225 & .839 &  & -.019 & .288 & .216 & .857 \\ 
  $\gamma_1$ & .003 & .161 & .158 & .946 &  & -.014 & .155 & .155 & .952 &  & -.019 & .159 & .156 & .949 &  & -.021 & .152 & .141 & .931 \\ 
  $\gamma_2$ & .001 & .265 & .269 & .955 &  & .019 & .255 & .261 & .958 &  & .024 & .264 & .266 & .959 &  & .026 & .251 & .208 & .889 \\ 
  $\sigma^2$ & .028 & .159 & .158 & .955 &  & -.172 & .049 & .112 & .765 &  & -.098 & .069 & .148 & .997 &  & -- & -- & -- & -- \\
  \multicolumn{20}{c}{\underline{Setting 3: $p=0.9$, $q=0.6$, $r=0$}} \\
  $\beta_1$ & .018 & .177 & .178 & .947 &  & -.300 & .125 & .122 & .316 &  & -.233 & .146 & .144 & .570 &  & -.039 & .160 & .151 & .932 \\ 
  $\beta_2$ & .022 & .342 & .328 & .942 &  & -.412 & .230 & .223 & .535 &  & -.267 & .254 & .257 & .826 &  & -.002 & .315 & .227 & .850 \\ 
  $\gamma_1$ & .006 & .162 & .159 & .947 &  & .023 & .169 & .167 & .942 &  & .000 & .170 & .162 & .946 &  & -.021 & .152 & .141 & .933 \\ 
  $\gamma_2$ & -.001 & .266 & .271 & .955 &  & -.019 & .277 & .279 & .956 &  & .005 & .277 & .276 & .955 &  & .026 & .251 & .212 & .896 \\ 
  $\sigma^2$ & .059 & .211 & .189 & .941 &  & .259 & .241 & .199 & .783 &  & .085 & .347 & .234 & .940 &  & -.250 & -- & -- & -- \\ 
  \multicolumn{20}{c}{\underline{Setting 4: $p=0.8$, $q=0.5$, $r=1$}} \\
  $\beta_1$ & .026 & .197 & .187 & .948 &  & -.415 & .094 & .097 & .010 &  & -.304 & .112 & .111 & .213 &  & -.019 & .177 & .157 & .917 \\ 
  $\beta_2$ & .037 & .365 & .354 & .947 &  & -.272 & .182 & .179 & .652 &  & -.245 & .206 & .207 & .773 &  & -.022 & .326 & .236 & .849 \\ 
  $\gamma_1$ & .003 & .161 & .158 & .945 &  & -.017 & .153 & .155 & .954 &  & -.022 & .158 & .155 & .947 &  & -.021 & .152 & .142 & .933 \\ 
  $\gamma_2$ & .001 & .266 & .270 & .955 &  & .023 & .253 & .266 & .960 &  & .029 & .263 & .265 & .957 &  & .026 & .251 & .217 & .912 \\ 
  $\sigma^2$ & .027 & .165 & .147 & .935 &  & -.211 & .012 & .103 & .342 &  & -.132 & .040 & .117 & 1.000 &  & -- & -- & -- & -- \\
  \multicolumn{20}{c}{\underline{Setting 5: $p=0.8$, $q=0.5$, $r=0.5$}} \\
  $\beta_1$ & .032 & .227 & .219 & .948 &  & -.401 & .096 & .098 & .020 &  & -.326 & .121 & .121 & .226 &  & -.016 & .204 & .182 & .915 \\ 
  $\beta_2$ & .045 & .429 & .413 & .944 &  & -.364 & .190 & .181 & .462 &  & -.292 & .226 & .224 & .736 &  & -.013 & .384 & .259 & .824 \\ 
  $\gamma_1$ & .004 & .162 & .158 & .944 &  & -.013 & .155 & .153 & .946 &  & -.018 & .160 & .156 & .949 &  & -.021 & .152 & .144 & .936 \\ 
  $\gamma_2$ & -.001 & .266 & .270 & .952 &  & .019 & .256 & .261 & .955 &  & .024 & .265 & .267 & .958 &  & .026 & .251 & .225 & .922 \\ 
  $\sigma^2$ & .042 & .195 & .176 & .940 &  & -.165 & .035 & .108 & .859 &  & -.095 & .059 & .145 & .999 &  & -- & -- & -- & -- \\ 
  \multicolumn{20}{c}{\underline{Setting 6: $p=0.8$, $q=0.5$, $r=0$}} \\
  $\beta_1$ & .038 & .285 & .272 & .951 &  & -.360 & .118 & .121 & .146 &  & -.333 & .140 & .139 & .321 &  & -.025 & .258 & .224 & .910 \\ 
  $\beta_2$ & .045 & .529 & .507 & .940 &  & -.490 & .225 & .223 & .414 &  & -.373 & .260 & .253 & .670 &  & .019 & .486 & .297 & .775 \\ 
  $\gamma_1$ & .008 & .165 & .159 & .939 &  & .019 & .167 & .163 & .943 &  & -.004 & .167 & .160 & .945 &  & -.021 & .152 & .145 & .937 \\ 
  $\gamma_2$ & -.003 & .267 & .272 & .958 &  & -.015 & .273 & .277 & .951 &  & .011 & .274 & .274 & .959 &  & .026 & .251 & .232 & .929 \\ 
  $\sigma^2$ & .081 & .330 & .227 & .895 &  & .219 & .197 & .164 & .767 &  & .049 & .251 & .197 & .916 &  & -- & -- & -- & -- \\ 
  \bottomrule
\end{tabular}
\end{table}

\section{Application to The ADNI Data} \label{sec:application}

In our application to the ADNI data, we controlled for six baseline covariates: age, sex, marital status, years of education, retirement status, and ApoE4 genotype. Additionally, we included two time-varying covariates, amyloid-beta (Abeta) and Tau, collected from biospecimen assessments and both standardized, as they serve as predictive biomarkers for Alzheimer's disease. Generally, Abeta levels fluctuate and decrease over time, while Tau levels steadily increase. The time-varying covariates were imputed using the last observation carried forward.

We present the estimated coefficients obtained from our method in Table \ref{tab:coef}. The algorithm requires fewer than 500 iterations to converge, and the computation time is about 3 hours. These coefficients are interpreted as the logarithms of hazard ratios. The results indicate that older individuals with the ApoE4 genotype mutation have a higher hazard of developing AD and experiencing mortality. Amyloid-beta and Tau are highly predictive of AD. The estimated standard deviation of the random effect is $\widehat\sigma=2.872$, suggesting a strong correlation between the risks of AD and mortality. Table \ref{tab:coef} also includes coefficient estimates obtained using the First-Diag and Last-Diag methods. All methods indicate that amyloid-beta is significantly negatively associated with the risk of AD, and Tau is significantly positively associated with the risk of both AD and death. While the competing methods suggest a negative effect of amyloid-beta on mortality, the coefficient is not statistically significant in all methods.

\begin{table}[!tbt]
\centering
\caption{Estimated regression coefficients in the ADNI application} \label{tab:coef}
\begin{tabular}{lclcccccc}
  \toprule
 & & & \multicolumn{3}{c}{Alzheimer's disease} & \multicolumn{3}{c}{Death} \\
 \cmidrule(lr){4-6} \cmidrule(lr){7-9}
 Method & & Covariate & Est & SE & $P$-value & Est & SE & $P$-value \\ 
  \midrule
 Propose method & & Age  & 0.136 & 0.040 & 0.001 & 0.145 & 0.031 & $<$0.001 \\ 
 & & Education  & -0.118 & 0.100 & 0.237 & 0.030 & 0.076 & 0.694 \\ 
  & &  Sex  & -0.586 & 0.617 & 0.342 & -1.041 & 0.439 & 0.018 \\ 
  & &  Marital status & 1.455 & 0.810 & 0.072 & -0.491 & 0.503 & 0.329 \\ 
  & &  Retirement  & 1.000 & 0.739 & 0.176 & 0.822 & 0.727 & 0.258 \\ 
  & &  ApoE4  & 1.251 & 0.528 & 0.018 & 1.032 & 0.432 & 0.017 \\ 
  & &  Abeta & -1.554 & 0.329 & $<$0.001 & 0.034 & 0.188 & 0.857 \\ 
 & &   Tau  & 1.105 & 0.217 & $<$0.001 & 0.566 & 0.207 & 0.006 \\ 
 & &   $\sigma^2$ & 8.246 & 0.904 & $<$0.001 &  &  &  \\  \\

First-Diag & &  Age  & 0.052 & 0.019 & 0.008 & 0.094 & 0.031 & 0.002 \\ 
  & &  Education  & -0.071 & 0.049 & 0.150 & 0.027 & 0.067 & 0.690 \\ 
 & &   Sex  & -0.264 & 0.313 & 0.399 & -1.256 & 0.447 & 0.005 \\ 
  & &  Marital status  & 0.660 & 0.432 & 0.127 & -0.966 & 0.566 & 0.088 \\ 
  & &  Retirement  & 0.411 & 0.362 & 0.257 & 0.937 & 0.563 & 0.096 \\ 
  & &  ApoE4   & 0.301 & 0.300 & 0.316 & 0.783 & 0.404 & 0.052 \\ 
  & &  Abeta  & -1.838 & 0.182 & $<$0.001 & -0.188 & 0.177 & 0.287 \\ 
  & &   Tau  & 0.841 & 0.133 & $<$0.001 & 0.378 & 0.159 & 0.018 \\
 & &   $\sigma^2$ & 3.191 & 0.470 & $<$0.001 &  &  &  \\  \\

Last-Diag & &   Age  & 0.067 & 0.019 & 0.001 & 0.107 & 0.031 & 0.001 \\
  & &  Education & -0.066 & 0.049 & 0.179 & -0.015 & 0.067 & 0.821 \\  
 & &   Sex  & -0.238 & 0.313 & 0.446 & -1.684 & 0.447 & $<$0.001 \\ 
 & &   Marital status  & 0.409 & 0.432 & 0.343 & -1.404 & 0.566 & 0.013 \\ 
 & &   Retirement  & 0.582 & 0.362 & 0.108 & 1.301 & 0.563 & 0.021 \\ 
 & &   ApoE4   & 0.237 & 0.300 & 0.429 & 0.848 & 0.404 & 0.036 \\ 
 & &   Abeta & -1.889 & 0.182 & $<$0.001 & -0.247 & 0.177 & 0.163 \\ 
 & &   Tau & 1.065 & 0.133 & $<$0.001 & 0.553 & 0.159 & 0.001 \\
 & &   $\sigma^2$ & 4.428 & 0.623 & $<$0.001 &  &  &  \\
\\

 NoRandEff & & Age & 0.055 & 0.023 & 0.018 & 0.078 & 0.023 & 0.001 \\ 
 & &  Education & -0.057 & 0.055 & 0.305 & 0.008 & 0.047 & 0.871 \\ 
 & &  Sex & -0.273 & 0.365 & 0.455 & -0.893 & 0.301 & 0.003 \\ 
 & &  Marital status & 0.882 & 0.499 & 0.077 & -0.746 & 0.386 & 0.053 \\ 
 & &  Retirement & 0.567 & 0.522 & 0.277 & 0.589 & 0.414 & 0.155 \\ 
 & &  ApoE4 & 0.684 & 0.315 & 0.030 & 0.612 & 0.265 & 0.021 \\ 
 & &  Abeta & -1.300 & 0.210 & $<$0.001 & -0.194 & 0.120 & 0.106 \\ 
 & &  Tau & 0.446 & 0.129 & 0.001 & 0.257 & 0.117 & 0.028 \\ 
\bottomrule
\end{tabular}
\end{table}

\begin{figure}[!tbh]
    \centering
    \includegraphics[width=0.8\textwidth]{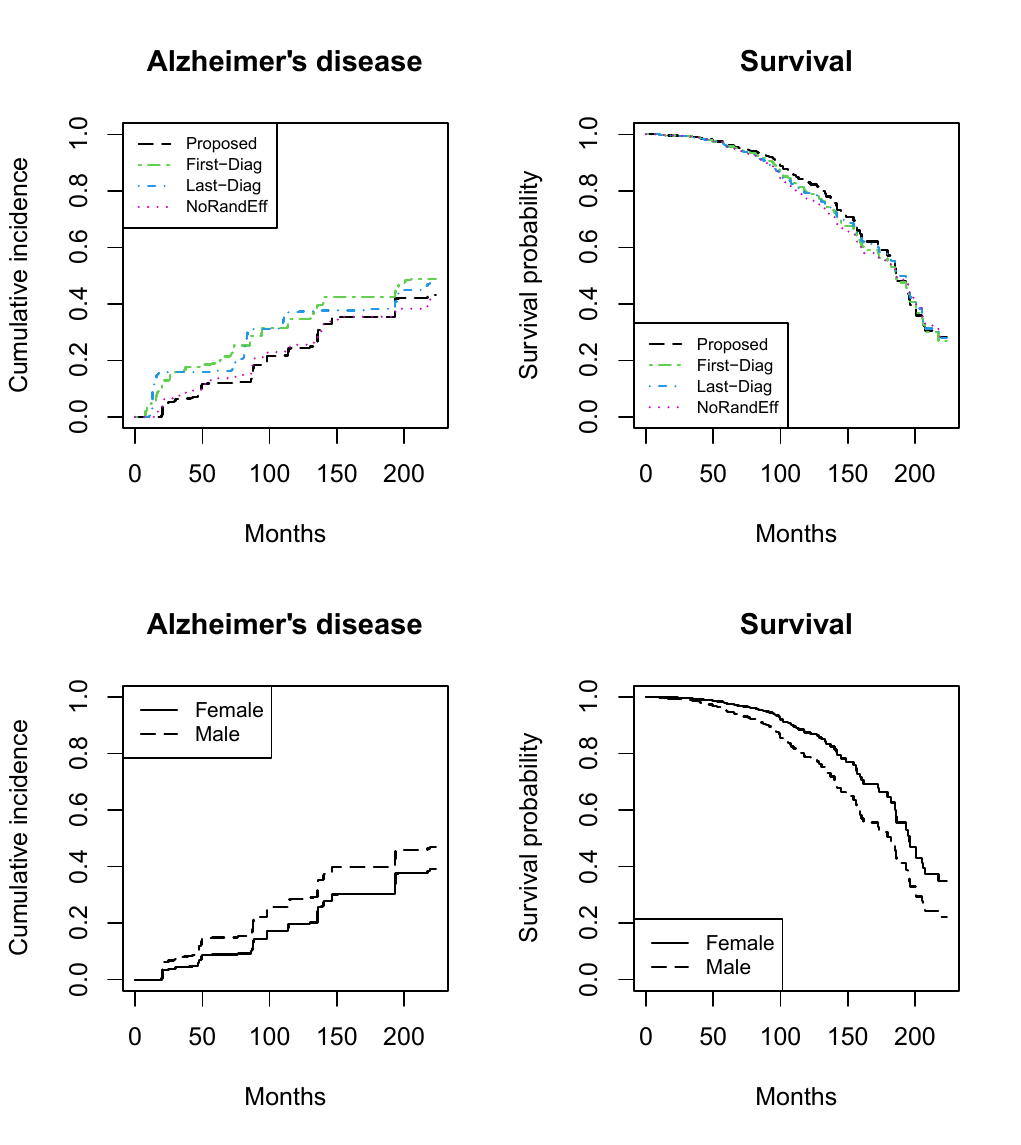}
    \caption{Estimated cumulative incidence function of Alzheimer's disease (AD) and death. Upper: estimated curves using different methods. Lower: estimated curves categorized by sex using the proposed method (47.4\% of the sample were female, and 52.6\% were male).} \label{surv_fit}
\end{figure}

The upper panels of Figure \ref{surv_fit} present the estimated cumulative incidence function of AD and the survival probability. The competing methods by First-Diag and Last-Diag tend to overestimate the risk of AD because the specificity of diagnosis is much lower than the sensitivity, which is also confirmed by the simulation study. Although many clinical diagnoses showed positive results, the true disease rate could be much lower. The method that ignores the random effect (NoRandEff) gives a similar estimate of the cumulative incidence function as the proposed method, but underestimates the survival probability. The lower panels of Figure \ref{surv_fit} show the estimated cumulative incidence function of AD and the survival probability by sex using the proposed method. Females have a higher survival probability than males.

\begin{figure}[!tbh]
    \centering
    \includegraphics[width=0.8\textwidth]{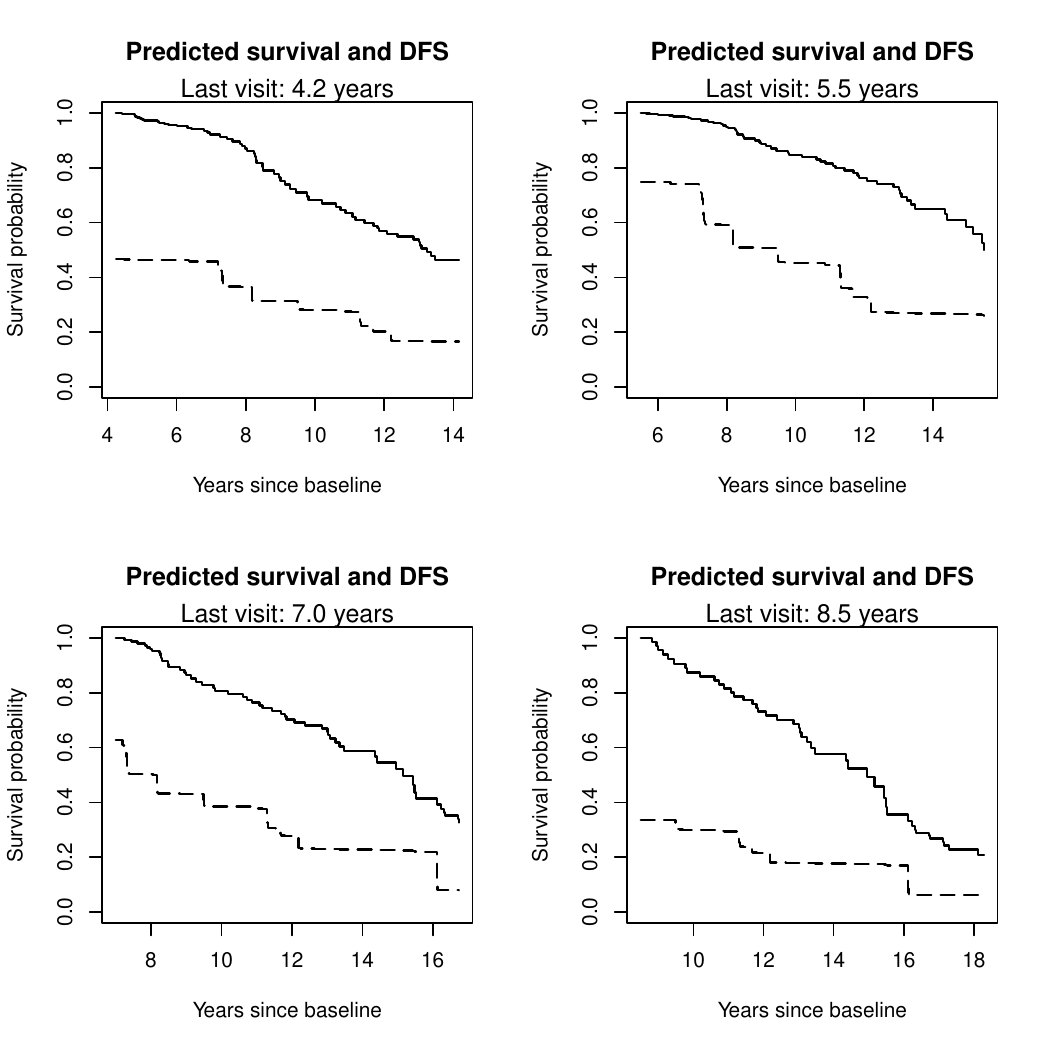}
    \caption{Estimated survival probability (in solid lines) and disease-free survival probability (in dashed lines) for an individual (ID=41) since the last visit.} \label{surv_predict}
\end{figure}

To demonstrate how to predict survival probabilities, we focus on an individual in the sample who was still alive at the last recorded visit (ID=41). This individual was a 71-year-old, retired, married female with 14 years of education and an ApoE4 genotype mutation. Her first three diagnoses were negative, while the last four were positive. Using the covariates and diagnosis information up to the final visit (4.2 years after baseline), we plotted the predicted survival and disease-free survival probabilities in the top-left panel of Figure \ref{surv_predict}.
If she received a negative diagnosis at 5.5 years post-baseline, the updated predicted survival curves are displayed in the top-right panel. A negative diagnosis during follow-up significantly decreases the likelihood of having AD, indicated by the dashed line in this panel. Further assuming she received two additional positive diagnoses at 7.0 and 8.5 years, the predicted curves for these visits are shown in the bottom-left and bottom-right panels, respectively. The predicted disease-free survival probabilities decline following positive diagnoses. Using follow-up clinical diagnoses increases the efficiency of prediction compared to only using baseline information.

\section{Discussion} \label{discussion}

Medical studies frequently encounter variability in diagnosis, which can lead to biased estimates of disease risks when relying on inaccurate diagnostic information. In the case of Alzheimer's disease, existing methods often overestimate these risks. In this article, we proposed a semiparametric modeling approach for a disease event subject to interval censoring and death subject to right censoring. The Cox proportional hazards model provided flexibility in specifying baseline hazards and offers straightforward interpretations for the coefficients. A shared random effect was utilized to capture the correlation between disease and death hazards. Misdiagnosis-induced mixtures of disease incidences resulted in challenges in identifying and estimating model parameters. We established the identifiability and demonstrated the asymptotic properties of the estimator. Our proposed estimation technique was made tractable by introducing missing data through an EM algorithm. While the estimated baseline hazard for the interval-censored event converges at a rate of $O_p(n^{-1/3})$, the estimator for the parametric component can converge at a rate of $O_p(n^{-1/2})$ and achieve asymptotic efficiency. 

Although diagnoses can be subject to measurement errors, they still offer valuable information about disease onset, as the posterior probabilities of the disease occurring in each interval are distinct. Higher sensitivity and specificity can help reduce the uncertainty in inferring the onset of the disease. In addition, to improve estimation accuracy, accurate diagnosis data can be utilized to provide more precise information about disease status. For example, postmortem examinations can be considered equivalent to a diagnosis at the time of death, with 100\% sensitivity and specificity. However, relying solely on postmortem data, without incorporating diagnostic information at previous visits, cannot identify the variance of the random effect, as autopsies are only conducted on deceased individuals and do not provide details on disease onset in survivors. The proposed model is useful for predicting survival and disease-free survival rates. New diagnostic information and measurements of time-varying covariates can update these predictions. By taking into account misdiagnoses, the dynamic prediction utilizes all available information, so both the prediction accuracy and efficiency can be improved. Patients identified as being at high risk of diseases should be monitored more frequently. 

There are several directions for future research. First, sensitivity and specificity may vary over time. Diagnoses made closer to death might be more accurate, as they can use more historical information. The proposed framework can be adapted to incorporate time-varying sensitivity and specificity. Second, inaccurate sensitivity and specificity can lead to biased estimations. When these values are unknown, Bayesian approaches can be used by assuming an informative prior distribution for both parameters, typically derived from existing studies. Third, the semiparametric modeling approach can be extended to account for multiple events. Clinical assessments may provide information on the underlying cause of dementia, such as Alzheimer's disease or vascular dementia, each with different diagnostic accuracies. Fourth, joint modeling of longitudinal covariates and time-to-event outcomes can be explored. In this context, the model for time-varying covariates can include random intercepts or slopes, reflecting latent factors that influence both the progression of biomarkers and the occurrence of events. To appropriately address the hierarchical correlation, multivariate random effects may be necessary in the joint model. However, model identification and estimation could become more challenging when multiple random effects are involved.

\section{Significance Statement}
AD is a progressive neurodegenerative disorder and is one of the leading causes of death worldwide. Based on the data analysis in the ADNI dataset, we found that amyloid-beta is significantly negatively associated with the risk of AD, and Tau is significantly positively associated with the risk of both AD and death. There is a strong correlation between the risks of AD and death. Our proposed method can facilitate the dynamic prediction of the risks of AD and death based on diagnostic data, which may be subject to misdiagnoses.

\begin{acks}[Acknowledgments]
The authors would like to thank the anonymous referees, an Associate Editor, and the Editor for their constructive comments that improved the quality of this paper.
\end{acks}

\begin{funding}
The second and third authors were partially supported by U.S. NIH Grants NS073671, MH123487, and GM124104.
\end{funding}

\begin{supplement}
\stitle{Supplementary material}
\sdescription{The online Supplementary Material includes (A) proof of Lemma \ref{lem1}, (B) proof of Theorem \ref{thm1}, (C) proof of Theorem \ref{thm2}, (D) useful lemmas, (E) additional simulation results, and (F) additional results in the real-data application.}
\end{supplement}

\bibliographystyle{imsart-nameyear}
\bibliography{ref}



\section*{Supplementary material}

# Supplementary Material for "Semiparametric Analysis of Interval-Censored Data Subject to Inaccurate Diagnoses with A Terminal Event"

Yuhao Deng, Donglin Zeng and Yuanjia Wang

## A  Proof of Lemma 1

We only consider the situation when postmortem data are not available, since having postmortem data is a special case. We set $\Delta_i = 1$ and $G_i = 0$. Suppose that the likelihood functions from $(\boldsymbol{\theta}_*, \mathcal{A}_*)$ and $(\boldsymbol{\theta}_0, \mathcal{A}_))$ are the same. For any fixed $l = 0, ..., N_i$, we consider all possible $\xi_{il*} \in \{0,1\}$ for $l^* = 0, ..., N_i$ except $l$, and sum over the equalities from the two likelihood functions. We obtain

$$\int_b \nu_*(Y_i) e^{\boldsymbol{\gamma}_*^\top \mathbf{X}_i(Y_i)+b} \exp\left\{-\int_0^{Y_i} e^{\boldsymbol{\gamma}_*^\top \mathbf{X}_i(s)+b} dV_*(s)\right\}$$

$$\times \left[q^{1-\xi_{il}}(1-q)^{\xi_{il}} \exp\left\{-\int_0^{Q_{il}} e^{\boldsymbol{\gamma}_*^\top \mathbf{X}_i(s)+b} d\Lambda_*(s)\right\}\right.$$

$$\left. - p^{\xi_{il}}(1-p)^{1-\xi_{il}} \exp\left\{-\int_0^{Q_{i,l+1}} e^{\boldsymbol{\gamma}_*^\top \mathbf{X}_i(s)+b} d\Lambda_*(s)\right\}\right] \psi(b; \sigma_*^2) db$$

$$= \int_b \nu_0(Y_i) e^{\boldsymbol{\gamma}_0^\top \mathbf{X}_i(Y_i)+b} \exp\left\{-\int_0^{Y_i} e^{\boldsymbol{\gamma}_0^\top \mathbf{X}_i(s)+b} dV_0(s)\right\}$$

$$\times \left[q^{1-\xi_{il}}(1-q)^{\xi_{il}} \exp\left\{-\int_0^{Q_{il}} e^{\boldsymbol{\gamma}_0^\top \mathbf{X}_i(s)+b} d\Lambda_0(s)\right\}\right.$$

$$\left. - p^{\xi_{il}}(1-p)^{1-\xi_{il}} \exp\left\{-\int_0^{Q_{i,l+1}} e^{\boldsymbol{\gamma}_0^\top \mathbf{X}_i(s)+b} d\Lambda_0(s)\right\}\right] \psi(b; \sigma_0^2) db.$$

This results in two equalities for $\xi_{il} = 0$ and $1$. According to Condition 5, $\begin{pmatrix} q & 1-q \\ 1-p & p \end{pmatrix}$ is non-singular so these two equalities yield

$$\int_b \nu_*(Y_i) e^{\boldsymbol{\gamma}_*^\top \mathbf{X}_i(Y_i)+b} \exp\left\{-\int_0^{Y_i} e^{\boldsymbol{\gamma}_*^\top \mathbf{X}_i(s)+b} dV_*(s)\right\}$$
$$\times \exp\left\{-\int_0^{Q_{il}} e^{\boldsymbol{\beta}_*^\top \mathbf{X}_i(s)+b} d\Lambda_*(s)\right\} \psi(b; \sigma_*^2) db \tag{S1}$$
$$= \int_b \nu_0(Y_i) e^{\boldsymbol{\gamma}_0^\top \mathbf{X}_i(Y_i)+b} \exp\left\{-\int_0^{Y_i} e^{\boldsymbol{\gamma}_0^\top \mathbf{X}_i(s)+b} dV_0(s)\right\}$$
$$\times \exp\left\{-\int_0^{Q_{il}} e^{\boldsymbol{\beta}_0^\top \mathbf{X}_i(s)+b} d\Lambda_0(s)\right\} \psi(b; \sigma_0^2) db.$$

1

We integrate $Y_i$ from 0 to $t$ to obtain

$$\int_b \exp\left\{-\int_0^t e^{\boldsymbol{\gamma}_*^\top \mathbf{X}_i(s)+b} d\mathrm{V}_*(s)\right\} \exp\left\{-\int_0^{Q_{il}} e^{\boldsymbol{\beta}_*^\top \mathbf{X}_i(s)+b} d\Lambda_*(s)\right\} \psi(b;\sigma_*^2) db \qquad (S2)$$
$$= \int_b \exp\left\{-\int_0^t e^{\boldsymbol{\gamma}_0^\top \mathbf{X}_i(s)+b} d\mathrm{V}_0(s)\right\} \exp\left\{-\int_0^{Q_{il}} e^{\boldsymbol{\beta}_0^\top \mathbf{X}_i(s)+b} d\Lambda_0(s)\right\} \psi(b;\sigma_0^2) db.$$

In particular, setting $l=0$, i.e., $Q_{il}=0$, in (S2) gives

$$\int_b \exp\left\{-\int_0^t e^{\boldsymbol{\gamma}_*^\top \mathbf{X}_i(s)+b} d\mathrm{V}_*(s)\right\} \psi(b;\sigma_*^2) db = \int_b \exp\left\{-\int_0^t e^{\boldsymbol{\gamma}_0^\top \mathbf{X}_i(s)+b} d\mathrm{V}_*(s)\right\} \psi(b;\sigma_0^2) db.$$

It follows from Elbers and Ridder (1982) that $\sigma_*^2 = \sigma_0^2$ and

$$\int_0^t e^{\boldsymbol{\gamma}_*^\top \mathbf{X}_i(s)} d\mathrm{V}_*(s) = \int_0^t e^{\boldsymbol{\gamma}_0^\top \mathbf{X}_i(s)} d\mathrm{V}_0(s).$$

We differentiate both sides with respect to $t$ and take the logarithm,

$$\boldsymbol{\gamma}_*^\top \mathbf{X}_i(t) + \log\{\nu_*(t)\} = \boldsymbol{\gamma}_0^\top \mathbf{X}_i(t) + \log\{\nu_0(t)\}.$$

By Condition 2, $\boldsymbol{\gamma}_* = \boldsymbol{\gamma}_0$ and $\mathrm{V}_*(t) = \mathrm{V}_0(t)$. On the other hand, we let $Y_i = 0$ and $Q_{il} = t$ in (S2). We obtain

$$\int_b \exp\left\{-\int_0^t e^{\boldsymbol{\beta}_*^\top \mathbf{X}_i(s)+b} d\Lambda_*(s)\right\} \psi(b;\sigma_0^2) db = \int_b \exp\left\{-\int_0^t e^{\boldsymbol{\beta}_0^\top \mathbf{X}_i(s)+b} d\Lambda_0(s)\right\} \psi(b;\sigma_0^2) db. \quad (S3)$$

Similarly, Elbers and Ridder (1982) and Condition 2 yield $\boldsymbol{\beta}_* = \boldsymbol{\beta}_0$ and $\Lambda_*(t) = \Lambda_0(t)$.

## B Proof of Theorem 2

Let $\boldsymbol{\theta} = (\boldsymbol{\beta}^\top, \boldsymbol{\gamma}^\top, \sigma^2)^\top$ and $\mathcal{A} = (\mathrm{V}(\cdot), \Lambda(\cdot))$ denote the parametric part and nonparametric (nuisance) part in the model, respectively. Let $l_n(\boldsymbol{\theta}, \mathcal{A})$ be the log-likelihood function. Let $\widetilde{M} = \sup_{t \in [0,\tau]} \sup_{\mathbf{X}_i(t)} |\boldsymbol{\gamma}^\top \mathbf{X}_i(t)|$. Since $\boldsymbol{\gamma}$ is in a compact set by Condition 1 and $X(\cdot)$ has bounded total variation by Condition 2, $\widetilde{M}$ is finite. For any $(\boldsymbol{\theta}, \mathcal{A})$ in the parameter space, the likelihood contributed by the $i$th unit is bounded by

$$\int_b \{\nu\{Y_i\} e^{\boldsymbol{\gamma}^\top \mathbf{X}_i(Y_i)+b}\}^{\Delta_i} \exp\left\{-\int_0^{Y_i} e^{\boldsymbol{\gamma}^\top \mathbf{X}_i(s)+b} d\mathrm{V}(s)\right\} \psi(b;\sigma^2) db$$
$$\leq \int_b \{\nu\{Y_i\} e^{\boldsymbol{\gamma}^\top \mathbf{X}_i(Y_i)+b}\}^{\Delta_i} \left\{1 + \int_0^{Y_i} e^{\boldsymbol{\gamma}^\top \mathbf{X}_i(s)+b} d\mathrm{V}(s)\right\}^{-1} \psi(b;\sigma^2) db$$
$$\leq \int_b \left\{\frac{\nu\{Y_i\} e^{\boldsymbol{\gamma}^\top \mathbf{X}_i(Y_i)+b}}{1 + \int_0^{Y_i} e^{\boldsymbol{\gamma}^\top \mathbf{X}_i(s)+b} d\mathrm{V}(s)}\right\}^{\Delta_i} \psi(b;\sigma^2) db$$
$$\leq \int_b \left\{\frac{\nu\{Y_i\} e^{\widetilde{M}} e^b}{1 + \mathrm{V}(Y_i) e^{-\widetilde{M}} e^b}\right\} \psi(b;\sigma^2) db \leq \int_b e^{2\widetilde{M}} \psi(b;\sigma^2) db = e^{2\widetilde{M}},$$

so the nonparametric likelihood function has a finite maximum. The estimator $(\widehat{\boldsymbol{\theta}}, \widehat{\mathcal{A}})$ that maximizes the nonparametric likelihood function exists. By definition, $l_n(\widehat{\boldsymbol{\theta}}, \widehat{\mathcal{A}}) - l_n(\boldsymbol{\theta}, \mathcal{A}) \geq 0$ for any $(\boldsymbol{\theta}, \mathcal{A})$ in the parameter space.

Next, we construct two step functions for $\Lambda$ and V, respectively, such that they converge to the true values. By Condition 4, the measure generated by $E\{\sum_{l=0}^{N+1} I(Q_{il}^* \leq t)\}$ is dominated by the sum of the Lebesgue measure on $[0, \tau]$ and a point measure at $\tau$, and its Radon-Nikodym derivative $f_1(t)$ is bounded away from zero. For the interval-censored event, we construct $\widetilde{\Lambda}(t)$ as a step function with jumps on $\mathcal{Q}$, i.e.,

$$\widetilde{\Lambda}(t) = \int_0^t \frac{\lambda_0(s)}{f_1(s)} d\left\{n^{-1} \sum_{i=1}^n \sum_{l=0}^{N_i+1} I(Q_{il}^* \leq s)\right\}.$$

Since

$$n^{-1} \sum_{i=1}^n \sum_{l=0}^{N_i+1} I(Q_{il}^* \leq t) \to E\left\{\sum_{l=0}^{N+1} I(Q_l^* \leq t)\right\}$$

uniformly on $\mathcal{Q}$ with probability 1, it follows that $\widetilde{\Lambda}(t) \to \Lambda(t)$ uniformly in $\mathcal{Q}$. For the right-censored event, we construct $\widetilde{V}(t)$ as a step function with jump size $\widetilde{\nu}\{Y_i\}$ on $\{Y_i : \Delta_i = 1, i = 1, \ldots, n\}$ by taking the derivative of the log-likelihood function $l_n(\boldsymbol{\theta}, \mathcal{A})$ with respect to $\nu\{t\}$ and setting the derivative at zero,

$$\frac{1}{\widetilde{\nu}\{t\}} \sum_{i=1}^n \Delta_i I(Y_i = t) - \sum_{i=1}^n H(Y_i, \mathbf{O}_i; \boldsymbol{\theta}_0, \mathcal{A}_0) = 0,$$

where

$$H(t, \mathbf{O}_i; \boldsymbol{\theta}, \mathcal{A}) = \frac{\int_b J_1(\mathbf{O}_i, b; \boldsymbol{\theta}, \mathcal{A}) J_2(t, \mathbf{O}_i, b; \boldsymbol{\theta}, \mathcal{A}) \psi(b; \sigma^2) db}{\int_b J_1(\mathbf{O}_i, b; \boldsymbol{\theta}, \mathcal{A}) \psi(b; \sigma^2) db}$$

with

$$J_1(\mathbf{O}_i, b; \boldsymbol{\theta}, \mathcal{A}) = \{e^{\boldsymbol{\gamma}^\top \mathbf{X}_i(Y_i)+b}\}^{\Delta_i} \exp\left\{-\int_0^{Y_i} e^{\boldsymbol{\gamma}^\top \mathbf{X}_i(s)+b} dV(s)\right\}$$

$$\times \left[\sum_{l=0}^{N_i} P(\boldsymbol{\xi}_i \mid T_i \in I_{il}) \left\{\exp\left(-\int_0^{Q_{il}^*} e^{\boldsymbol{\beta}^\top \mathbf{X}_i(s)+b} d\Lambda(s)\right) \right.\right.$$

$$\left.\left. - I(Q_{i,l+1}^* < \infty) \exp\left(-\int_0^{Q_{i,l+1}^*} e^{\boldsymbol{\beta}^\top \mathbf{X}_i(s)+b} d\Lambda(s)\right)\right\}\right]^{1-\Delta_i G_i(1-W_i)}$$

$$\times \left[P(\boldsymbol{\xi}_i \mid T_i \in I_{i,N_i}) \exp\left\{-\int_0^{Y_i} e^{\boldsymbol{\beta}^\top \mathbf{X}_i(s)+b} d\Lambda(s)\right\}\right]^{\Delta_i G_i(1-W_i)}$$

and

$$J_2(t; \mathbf{O}_i, b; \boldsymbol{\theta}, \mathcal{A}) = I(Y_i \geq t) e^{\boldsymbol{\gamma}^\top \mathbf{X}_i(t)+b}.$$

Let $\mathcal{D}_M$ be the class of increasing functions $f(\cdot)$ subject to $f(0) = 0$ and $f(\tau) < M$. By Lemma 1 of Gao et al. (2019) and the preservation property, the function classes

$$\mathcal{H} = \{H(t, \mathbf{O}; \boldsymbol{\theta}, \mathcal{A}) : t \in [0, \tau], \boldsymbol{\theta} \in \Theta, \mathcal{A} \in \mathcal{D}_M \times \mathcal{D}_M\}$$

and $\{\Delta I(Y \leq t) : t \in [0, \tau]\}$ are Glivenko-Cantelli, so

$$\widetilde{V}(t) = \int_0^t \frac{\mathbb{P}_n\{\Delta dI(Y \leq s)\}}{\mathbb{P}_n H(s, \mathbf{O}; \boldsymbol{\theta}_0, \mathcal{A}_0)} \to \int_0^t \frac{dP(Y \leq s, \Delta = 1)}{P(Y \geq s)} = V_0(t)$$

uniformly in $t \in [0, \tau]$. Denote $\widetilde{\mathcal{A}} = (\widetilde{V}(\cdot), \widetilde{\Lambda}(\cdot))$, so $\widetilde{\mathcal{A}} \to \mathcal{A}_0$ uniformly in $[0, \tau] \times \mathcal{Q}$.

3

By Lemma S1 in the supplementary material, $\widehat{V}(\tau)$ and $\widehat{\Lambda}(\tau - \epsilon)$ are bounded. By the Helly's selection lemma, every sequence of $\widehat{V}(\cdot)$ and $\widehat{\Lambda}(\cdot)$ has a subsequence pointwise converging to some functions $V_*(\cdot)$ and $\Lambda_*(\cdot)$, and $\widehat{\boldsymbol{\theta}}$ converges to some vector $\boldsymbol{\theta}_*$ in probability. Denote $\mathcal{A}_* = (V_*(\cdot), \Lambda_*(\cdot))$. Recall that

$$\widehat{V}(t) = \int_0^t \frac{\mathbb{P}_n\{\Delta dI(Y \leq s)\}}{\mathbb{P}_n H(s, \mathbf{O}; \widehat{\boldsymbol{\theta}}, \widehat{\mathcal{A}})}, \quad \widetilde{V}(t) = \int_0^t \frac{\mathbb{P}_n\{\Delta dI(Y \leq s)\}}{\mathbb{P}_n H(s, \mathbf{O}; \boldsymbol{\theta}_0, \mathcal{A}_0)},$$

and $\mathcal{H}$ is Glivenko-Cantelli, so

$$\widehat{V}(t) = \int_0^t \frac{\mathbb{P}_n H(s, \mathbf{O}; \boldsymbol{\theta}_0, \mathcal{A}_0)}{\mathbb{P}_n H(s, \mathbf{O}; \widehat{\boldsymbol{\theta}}, \widehat{\mathcal{A}})} d\widetilde{V}(s) = \int_0^t \frac{\mathbb{P}H(s, \mathbf{O}; \boldsymbol{\theta}_0, \mathcal{A}_0)}{\mathbb{P}H(s, \mathbf{O}; \widehat{\boldsymbol{\theta}}, \widehat{\mathcal{A}})} d\widetilde{V}(s) + o_p(1).$$

By the continuous mapping theorem, $\mathbb{P}H(t, \mathbf{O}; \widehat{\boldsymbol{\theta}}, \widehat{\mathcal{A}})$ converges to $\mathbb{P}H(t, \mathbf{O}; \boldsymbol{\theta}_*, \mathcal{A}_*)$ uniformly in $[0, \tau]$ along a subsequence with probability 1. Therefore,

$$\frac{\widehat{\nu}\{t\}}{\widetilde{\nu}\{t\}} = \frac{d\widehat{V}(t)}{d\widetilde{V}(t)} = \frac{\mathbb{P}H(t, \mathbf{O}; \boldsymbol{\theta}_0, \mathcal{A}_0)}{\mathbb{P}H(t, \mathbf{O}; \widehat{\boldsymbol{\theta}}, \widehat{\mathcal{A}})} \to \frac{\mathbb{P}H(t, \mathbf{O}; \boldsymbol{\theta}_0, \mathcal{A}_0)}{\mathbb{P}H(t, \mathbf{O}; \boldsymbol{\theta}_*, \mathcal{A}_*)} = \frac{dV_*(t)}{dV_0(t)} = \frac{\nu_*(t)}{\nu_0(t)}$$

uniformly in $[0, \tau]$ with probability 1. The ratio of jump sizes of $\widehat{V}(\cdot)$ and $\widetilde{V}(\cdot)$ converges to the ratio of derivatives of $V_*(\cdot)$ and $V_0(\cdot)$. The ratio is positive because $\mathbb{P}H(\cdot, \mathbf{O}; \boldsymbol{\theta}, \mathcal{A}) > 0$ and finite because

$$V_*(\tau) = \int_0^\tau \frac{\mathbb{P}H(s, \mathbf{O}; \boldsymbol{\theta}_0, \mathcal{A}_0)}{\mathbb{P}H(s, \mathbf{O}; \boldsymbol{\theta}_*, \mathcal{A}_*)} dV_0(s) \leq \limsup_n \widehat{V}(\tau) < \infty.$$

We now show that $\mathcal{A}_* = \mathcal{A}_0$ and $\boldsymbol{\theta}_* = \boldsymbol{\theta}_0$. Let

$$L(\boldsymbol{\theta}, \mathcal{A}) = \exp\{l(\boldsymbol{\theta}, \mathcal{A})\} = \int_b \nu\{Y\}^\Delta J_1(\mathbf{O}, b; \boldsymbol{\theta}, \mathcal{A}) \psi(b; \sigma^2) db$$

be the nonparametric likelihood function contributed by a single unit. Define

$$m(\boldsymbol{\theta}, \mathcal{A}) = \log\left\{\frac{L(\boldsymbol{\theta}, \mathcal{A}) + L(\boldsymbol{\theta}_0, \widetilde{\mathcal{A}})}{2}\right\}.$$

Since $\log(\cdot)$ is concave,

$$\mathbb{P}_n m(\widehat{\boldsymbol{\theta}}, \widehat{\mathcal{A}}) \geq \frac{1}{2}\left\{\mathbb{P}_n \log L(\widehat{\boldsymbol{\theta}}, \widehat{\mathcal{A}}) + \mathbb{P}_n \log L(\boldsymbol{\theta}_0, \widetilde{\mathcal{A}})\right\}$$

$$\geq \mathbb{P}_n \log L(\boldsymbol{\theta}_0, \widetilde{\mathcal{A}}) = \mathbb{P}_n m(\boldsymbol{\theta}_0, \widetilde{\mathcal{A}})$$

By Lemma 1 of Gao et al. (2019), $\mathcal{M} = \{m(\boldsymbol{\theta}, \mathcal{A}) : \boldsymbol{\theta} \in \Theta, \mathcal{A} \in \mathcal{D}_M \times \mathcal{D}_M\}$ is Glivenko-Cantelli, so $\mathbb{P}_n m(\widehat{\boldsymbol{\theta}}, \widehat{\mathcal{A}}) \to \mathbb{P}m(\widehat{\boldsymbol{\theta}}, \widehat{\mathcal{A}})$ and $\mathbb{P}_n m(\boldsymbol{\theta}_0, \widetilde{\mathcal{A}}) \to \mathbb{P}m(\boldsymbol{\theta}_0, \widetilde{\mathcal{A}})$ almost surely, and hence

$$O_p(1) \leq \mathbb{P}\left\{m(\widehat{\boldsymbol{\theta}}, \widehat{\mathcal{A}}) - m(\boldsymbol{\theta}_0, \widetilde{\mathcal{A}})\right\}$$

$$= \mathbb{P}\log\left\{\frac{1}{2} + \frac{\int_b \nu_*(Y)^\Delta J_1(\mathbf{O}, b; \boldsymbol{\theta}_*, \mathcal{A}_*)\psi(b; \sigma_*^2)db}{2\int_b \nu_0(Y)^\Delta J_1(\mathbf{O}, b; \boldsymbol{\theta}_0, \mathcal{A}_0)\psi(b; \sigma_0^2)db}\right\} + o_p(1)$$

$$= \mathbb{P}\log\left[\frac{\{p(\boldsymbol{\theta}_0, \mathcal{A}_0) + p(\boldsymbol{\theta}_*, \mathcal{A}_*)\}/2}{p(\boldsymbol{\theta}_0, \mathcal{A}_0)}\right] + o_p(1),$$

where $p(\boldsymbol{\theta}, \mathcal{A}) = \int_b \nu(t)^\Delta J_1(\mathbf{O}, b; \boldsymbol{\theta}, \mathcal{A})\psi(b; \sigma^2) db$ is the density function evaluated at $(\boldsymbol{\theta}, \mathcal{A})$. The inequality above implies that the negative Kullback–Leibler distance between $\{p(\boldsymbol{\theta}_0, \mathcal{A}_0) + p(\boldsymbol{\theta}_*, \mathcal{A}_*)\}/2$ and $p(\boldsymbol{\theta}, \mathcal{A})$ is positive, indicating that $p(\boldsymbol{\theta}_*, \mathcal{A}_*) = p(\boldsymbol{\theta}_0, \mathcal{A}_0)$ with probability 1. From Lemma 1, we conclude that $\boldsymbol{\theta}_* = \boldsymbol{\theta}_0$ and $\mathcal{A}_* = \mathcal{A}_0$ almost surely in $[0, \tau] \times \mathcal{Q}$. This implies that $\|\widehat{\boldsymbol{\theta}} - \boldsymbol{\theta}_0\| \to 0$ and $\|\widehat{\mathcal{A}} - \mathcal{A}\|_\infty \to 0$ on $[0, \tau] \times \mathcal{Q}$ with probability 1.

# C   Proof of Theorem 3

The proof is similar to Gao et al. (2019), so we only provide a sketched proof below. Let

$$H_2(t, \mathbf{O}_i; \boldsymbol{\theta}, \mathcal{A}) = \frac{\int_b \{J_3(t, \mathbf{O}_i, b; \boldsymbol{\theta}, \mathcal{A}) - J_4(t, \mathbf{O}_i, b; \boldsymbol{\theta}, \mathcal{A})\} \psi(b; \sigma^2) db}{\int_b J_1(\mathbf{O}_i, b; \boldsymbol{\theta}, \mathcal{A}) \psi(b; \sigma^2) db}$$

with

$$J_3(t, \mathbf{O}_i, b; \boldsymbol{\theta}, \mathcal{A}) = \{e^{\boldsymbol{\gamma}^\top \mathbf{X}_i(Y_i)+b}\}^{\Delta_i} \exp\left\{-\int_0^{Y_i} e^{\boldsymbol{\gamma}^\top \mathbf{X}_i(s)+b} d\mathrm{V}(s)\right\} e^{\boldsymbol{\beta}^\top \mathbf{X}_i(t)+b}$$

$$\times \left[\sum_{l=0}^{N_i} P(\boldsymbol{\xi}_i \mid T_i \in I_{il}) \exp\left\{-\int_0^{Q_{il}^*} e^{\boldsymbol{\beta}^\top \mathbf{X}_i(s)+b} d\Lambda(s)\right\} I(t \le Q_{il}^*)\right]^{1-\Delta_i G_i(1-W_i)}$$

$$\times \left[P(\boldsymbol{\xi}_i \mid T_i \in I_{i,N_i}) \exp\left\{-\int_0^{Y_i} e^{\boldsymbol{\beta}^\top \mathbf{X}_i(s)+b} d\Lambda(s)\right\} I(t \le Y_i)\right]^{\Delta_i G_i(1-W_i)},$$

$$J_4(t, \mathbf{O}_i, b; \boldsymbol{\theta}, \mathcal{A}) = \{e^{\boldsymbol{\gamma}^\top \mathbf{X}_i(Y_i)+b}\}^{\Delta_i} \exp\left\{-\int_0^{Y_i} e^{\boldsymbol{\gamma}^\top \mathbf{X}_i(s)+b} d\mathrm{V}(s)\right\} e^{\boldsymbol{\beta}^\top \mathbf{X}_i(t)+b}$$

$$\times \left[\sum_{l=0}^{N_i} P(\boldsymbol{\xi}_i \mid T_i \in I_{il}) \exp\left\{-\int_0^{Q_{i,l+1}^*} e^{\boldsymbol{\beta}^\top \mathbf{X}_i(s)+b} d\Lambda(s)\right\} I(t \le Q_{i,l+1}^*)\right]^{1-\Delta_i G_i(1-W_i)}.$$

Consider the submodel $d\mathrm{V}_\epsilon(\cdot) = \{1+\epsilon h_1(\cdot)\} d\mathrm{V}_0(\cdot)$ and $d\Lambda_\epsilon(\cdot) = \{1+\epsilon h_2(\cdot)\} d\Lambda_0(\cdot)$. The score functions along this submodel are

$$\ell_{\boldsymbol{\gamma}}(\boldsymbol{\theta}, \mathcal{A}) = \Delta X(Y) - \int_0^\tau H(t, \mathbf{O}; \boldsymbol{\theta}, \mathcal{A}) X(t) d\mathrm{V}(t),$$

$$\ell_{\boldsymbol{\beta}}(\boldsymbol{\theta}, \mathcal{A}) = \int_0^\tau H_2(t, \mathbf{O}; \boldsymbol{\theta}, \mathcal{A}) X(t) d\Lambda(t),$$

$$\ell_{\sigma^2}(\boldsymbol{\theta}, \mathcal{A}) = \frac{1}{\int_b J_1(\mathbf{O}; \boldsymbol{\theta}, \mathcal{A}) \psi(b; \sigma^2) db} \int_b J_1(\mathbf{O}; \boldsymbol{\theta}, \mathcal{A}) \left\{-\frac{1}{2\sigma^2} + \frac{b^2}{2\sigma^4}\right\} \psi(b; \sigma^2) db,$$

$$\ell_\mathrm{V}(\boldsymbol{\theta}, \mathcal{A})(h_1) = \Delta h_1(Y) - \int_0^\tau H(t, \mathbf{O}; \boldsymbol{\theta}, \mathcal{A}) h_1(t) d\mathrm{V}(t),$$

$$\ell_\Lambda(\boldsymbol{\theta}, \mathcal{A})(h_2) = \int_0^\tau H_2(t, \mathbf{O}; \boldsymbol{\theta}, \mathcal{A}) h_2(t) d\Lambda(t).$$

We denote $\ell_{\boldsymbol{\theta}}(\boldsymbol{\theta}, \mathcal{A}) = (\ell_{\boldsymbol{\beta}}, \ell_{\boldsymbol{\gamma}}, \ell_{\sigma^2})(\boldsymbol{\theta}, \mathcal{A})$ and $\ell_\mathcal{A}(\boldsymbol{\theta}, \mathcal{A})(h) = \ell_\mathrm{V}(\boldsymbol{\theta}, \mathcal{A})(h_1) + \ell_\Lambda(\boldsymbol{\theta}, \mathcal{A})(h_2)$. Denote $\mathbb{G}_n = \sqrt{n}(\mathbb{P}_n - \mathbb{P})$ as the empirical process. Then if letting $\ell_{\boldsymbol{\theta}\boldsymbol{\theta}}$, $\ell_{\boldsymbol{\theta}\mathcal{A}}(h)$, $\ell_{\mathcal{A}\mathcal{A}}(h, h^*)$ be the derivative of the score function evaluated at $(\boldsymbol{\theta}_0, \mathcal{A}_0)$, we obtain

$$\mathbb{G}_n \ell_{\boldsymbol{\theta}}(\widehat{\boldsymbol{\theta}}, \widehat{\mathcal{A}}) = -\sqrt{n}\mathbb{P}\{\ell_{\boldsymbol{\theta}\boldsymbol{\theta}}(\widehat{\boldsymbol{\theta}} - \boldsymbol{\theta}_0) + \ell_{\boldsymbol{\theta}\mathcal{A}}(\widehat{\mathcal{A}} - \mathcal{A}_0)\} + r_1,$$

$$\mathbb{G}_n \ell_\mathcal{A}(\widehat{\boldsymbol{\theta}}, \widehat{\mathcal{A}})(h) = -\sqrt{n}\mathbb{P}\{\ell_{\boldsymbol{\theta}\mathcal{A}}(\widehat{\boldsymbol{\theta}} - \boldsymbol{\theta}_0)(h) + \ell_{\mathcal{A}\mathcal{A}}(\widehat{\mathcal{A}} - \mathcal{A}_0, h)\} + r_2.$$

By Lemma S4 in the supplementary material, the second-order terms $r_1$ and $r_2$ are bounded by $O_p(\sqrt{n}\|\widehat{\boldsymbol{\theta}} - \boldsymbol{\theta}_0\|^2 + n^{-1/6})$.

We choose $h$ to be the least favorable direction $\mathbf{h}^*(t) = (\mathbf{h}_1^*(t)^\top, \mathbf{h}_2^*(t)^\top)^\top$, a vector with components in $L_2[0, \tau] \times L_2(\mathcal{Q})$ that solves the normal equation $\mathbb{P}\ell_{\mathcal{A}\mathcal{A}}(h, \mathbf{h}^*) = \mathbb{P}\ell_{\boldsymbol{\theta}\mathcal{A}}(h)$. The existence of $\mathbf{h}^*$ as well as

5

its continuous differentiability in $[0, \tau]$ can be shown following the similar arguments in Gao et al. (2019), except that we need to verify the one-to-one mapping property of the information operator for $\mathcal{A}$. The detail for the latter is given in Lemma S2. With these result, it holds

$$\mathbb{G}_n \ell_{\boldsymbol{\theta}}(\boldsymbol{\theta}_0, \mathcal{A}_0) = -\sqrt{n}\mathbb{P}\{\ell_{\boldsymbol{\theta\theta}}(\widehat{\boldsymbol{\theta}} - \boldsymbol{\theta}_0) + \ell_{\boldsymbol{\theta}\mathcal{A}}(\widehat{\mathcal{A}} - \mathcal{A}_0)\} + O_p(1),$$

$$\mathbb{G}_n \ell_{\mathcal{A}}(\boldsymbol{\theta}_0, \mathcal{A}_0)(\mathbf{h}^*) = -\sqrt{n}\mathbb{P}\{\ell_{\boldsymbol{\theta}\mathcal{A}}(\widehat{\boldsymbol{\theta}} - \boldsymbol{\theta}_0)(\mathbf{h}^*) + \ell_{\boldsymbol{\theta}\mathcal{A}}(\widehat{\mathcal{A}} - \mathcal{A}_0)\} + O_p(1).$$

Taking the difference for the above equations, we obtain

$$\mathbb{G}_n\{\ell_{\boldsymbol{\theta}}(\boldsymbol{\theta}_0, \mathcal{A}_0) - \ell_{\mathcal{A}}(\boldsymbol{\theta}_0, \mathcal{A}_0)(\mathbf{h}^*)\} = \sqrt{n}\mathbb{P}[\ell_{\boldsymbol{\theta}} - \ell_{\mathcal{A}}(\mathbf{h}^*)]^{\otimes 2}(\widehat{\boldsymbol{\theta}} - \boldsymbol{\theta}_0) + O_p(1).$$

We need $\mathbb{P}[\ell_{\boldsymbol{\theta}} - \ell_{\mathcal{A}}(\mathbf{h}^*)]^{\otimes 2}$ to be invertible. To see this, if $\mathbf{v}^\top \mathbb{P}[\ell_{\boldsymbol{\theta}} - \ell_{\mathcal{A}}(\mathbf{h}^*)]^{\otimes 2} \mathbf{v} = 0$ for some vector $\mathbf{v}$, then the score function $\ell(\boldsymbol{\theta}_0, \mathcal{A}_0) = \mathbf{v}^\top \ell_{\boldsymbol{\theta}}(\boldsymbol{\theta}_0, \mathcal{A}_0) - \mathbf{v}^\top \ell_{\mathcal{A}}(\boldsymbol{\theta}_0, \mathcal{A}_0)(\mathbf{h}^*)$ is zero with probability 1. By Lemma S3 in the supplementary material, $\mathbf{v} = 0$. Therefore, $\mathbb{P}[\ell_{\boldsymbol{\theta}} - \ell_{\mathcal{A}}(\mathbf{h}^*)]^{\otimes 2}$ is invertible, and

$$\sqrt{n}(\widehat{\boldsymbol{\theta}} - \boldsymbol{\theta}_0) = \{\mathbb{P}[\ell_{\boldsymbol{\theta}} - \ell_{\mathcal{A}}(\mathbf{h}^*)]^{\otimes 2}\}^{-1}\mathbb{G}_n\{\ell_{\boldsymbol{\theta}}(\boldsymbol{\theta}_0, \mathcal{A}_0) - \ell_{\mathcal{A}}(\boldsymbol{\theta}_0, \mathcal{A}_0)(\mathbf{h}^*)\} + O_p(1).$$

The influence function $\varphi = \{\mathbb{P}[\ell_{\boldsymbol{\theta}} - \ell_{\mathcal{A}}(\mathbf{h}^*)]^{\otimes 2}\}^{-1}\{\ell_{\boldsymbol{\theta}}(\boldsymbol{\theta}_0, \mathcal{A}_0) - \ell_{\mathcal{A}}(\boldsymbol{\theta}_0, \mathcal{A}_0)(\mathbf{h}^*)\}$ is in the tangent space spanned by score functions, so the asymptotic variance of $\widehat{\boldsymbol{\theta}}$ attains the semiparametric efficiency bound.

## D  Useful Lemmas

**Lemma S1.** *The estimated cumulative baseline hazards $\widehat{V}(\tau)$ and $\widehat{\Lambda}(\tau - \epsilon)$ are bounded for any $\epsilon > 0$.*

**Proof** Clearly, $n^{-1}\{l_n(\widehat{\boldsymbol{\theta}}, \widehat{\mathcal{A}}) - l_n(\boldsymbol{\theta}_0, \widetilde{\mathcal{A}})\} \geq 0$ by the definition of NPMLE. To see $\widehat{V}(\tau)$ is bounded, we note that

$$0 \leq n^{-1}\{l_n(\widehat{\boldsymbol{\theta}}, \widehat{\mathcal{A}}) - l_n(\boldsymbol{\theta}_0, \widetilde{\mathcal{A}})\}$$

$$\leq O(1) + n^{-1}\sum_{i=1}^n \log \int_b [\widehat{\nu}\{Y_i\}e^{\boldsymbol{\gamma}^\top X_i(Y_i)+b}]^{\Delta_i} \exp\left\{-\int_0^{Y_i} e^{\boldsymbol{\gamma}^\top X_i(s)+b}d\mathrm{V}(s)\right\}\psi(b; \widehat{\sigma}^2)db$$

$$\leq O(1) + n^{-1}\sum_{i=1}^n \Delta_i \log(\widehat{\nu}\{Y_i\}) + n^{-1}\sum_{i=1}^n \log \int_b \left\{\frac{e^{\widehat{\boldsymbol{\gamma}}^\top X_i(Y_i)+b}}{1 + \int_0^{Y_i} e^{\widehat{\boldsymbol{\gamma}}^\top X_i(s)+b}d\widehat{\mathrm{V}}(s)}\right\}\psi(b; \widehat{\sigma}^2)db$$

$$\leq O(1) + n^{-1}\sum_{i=1}^n \Delta_i \log(\widehat{\nu}\{Y_i\}) + n^{-1}\sum_{i=1}^n \log \int_b \left\{\frac{e^{\widetilde{M}+b}}{e^{-\widetilde{M}+b}\widehat{\mathrm{V}}(Y_i)}\right\}\psi(b; \widehat{\sigma}^2)db$$

$$\leq O(1) + n^{-1}\sum_{i=1}^n \Delta_i \log(\widehat{\nu}\{Y_i\}) - n^{-1}\sum_{i=1}^n \log(\widehat{\mathrm{V}}(Y_i)).$$

Let $0 = u_0 < u_1 < \cdots < u_K = \tau$ be a sequence of time points, so

$$n^{-1}\sum_{i=1}^n \Delta_i \log(\widehat{\nu}\{Y_i\}) - n^{-1}\sum_{i=1}^n \log(\widehat{\mathrm{V}}(Y_i))$$

$$\leq n^{-1}\sum_{i=1}^n \Delta_i \left\{\sum_{k=1}^K I(s_{k-1} \leq Y_i < s_k)\log(\widehat{\mathrm{V}}(s_k)) + I(Y_i = \tau)\log(\widehat{\mathrm{V}}(\tau))\right\}$$

$$- n^{-1}\sum_{i=1}^n \left\{\sum_{k=0}^{K-1} I(s_k \leq Y_i < s_{k+1})\log(\widehat{\mathrm{V}}(s_k)) + I(Y_i = \tau)\log(\widehat{\mathrm{V}}(\tau))\right\}$$

6

$$\leq -n^{-1} \sum_{i=1}^{n} \sum_{k=1}^{K-1} \{I(s_k \leq Y_i < s_{k+1} - \Delta_i I(s_{k-1} \leq Y_i < s_k))\} \log(\widehat{V}(s_k))$$

$$- n^{-1} \sum_{i=1}^{n} \{I(Y_i = \tau) - \Delta_i I(s_{K-1} \leq Y_i \leq \tau)\} \log(\widehat{V}(\tau)).$$

When $n$ is large, by Conditions 1 and 3, we can choose the sequence $\{s_k : k = 0, \ldots, K\}$ so that $P(s_k \leq Y_i < s_{k+1}) > P(s_k \leq Y_i < s_{k+1}, \Delta_i = 1)$ for each $k = 1, \ldots, K-1$ and $P(Y_i = \tau) > P(s_{K-1} \leq Y_i \leq \tau, \Delta_i = 1)$. Therefore, the coefficients of $\log(\widehat{V}(s_k))$ are all negative. If $\log(\widehat{V}(s_k))$ diverges to $\infty$, then the right-hand-side will diverge to $-\infty$, which contradicts with the fact that it should be non-negative.

If $\Lambda(\tau) < M$ in Condition 1, we can restrict the estimate $\widehat{\Lambda}(\cdot)$ to be smaller than $M$ in the NPMLE, so $\widehat{\Lambda}(\cdot)$ is bounded by $M$. If postmortem data are available, we can show $\widehat{\Lambda}(\tau - \epsilon)$ is bounded for any $\epsilon > 0$. To see this,

$$0 \leq n^{-1}\{l_n(\widehat{\boldsymbol{\theta}}, \widehat{\mathcal{A}}) - l_n(\boldsymbol{\theta}_0, \widetilde{\mathcal{A}})\}$$

$$\leq O(1) + n^{-1} \sum_{i=1}^{n} I\{\Delta_i G_i(1 - W_i) = 1, Q^*_{i,N_i+1} > \tau - \epsilon\}$$

$$\times \log \int_b e^{2\widetilde{M}} \left[ P(S_i = N_i \mid \boldsymbol{\xi}_i) \exp\left\{ -\int_0^{Q^*_{i,N_i+1}} e^{\widehat{\boldsymbol{\beta}}^\top X_i(s) + b} d\widehat{\Lambda}(s) \right\} \right] \psi(b; \widehat{\sigma}^2) db$$

$$\leq O(1) + n^{-1} \sum_{i=1}^{n} I\{\Delta_i G_i(1 - W_i) = 1, Q^*_{i,N_i+1} > \tau - \epsilon\}$$

$$\times \log \int_b \exp\left\{ -e^{-\widetilde{M}+b} \widehat{\Lambda}(\tau - \epsilon) \right\} \psi(b; \widehat{\sigma}^2) db.$$

For any constant $\delta > 0$ that is close to zero, we can find $B < 0$ small enough so that

$$\int_{b<B} \exp\left\{ -e^{-\widetilde{M}+b} \widehat{\Lambda}(\tau - \epsilon) \right\} \psi(b; \widehat{\sigma}^2) db \leq \int_{b<B} \psi(b; \widehat{\sigma}^2) db < \frac{\delta}{2}.$$

If $\limsup_n \widehat{\Lambda}(\tau - \epsilon) = \infty$, then

$$\int_{b \geq B} \exp\left\{ -e^{-\widetilde{M}+b} \widehat{\Lambda}(\tau - \epsilon) \right\} \psi(b; \widehat{\sigma}^2) db$$

$$\leq \int_{b \geq B} \exp\left\{ -e^{-\widetilde{M}+B} \widehat{\Lambda}(\tau - \epsilon) \right\} \psi(b; \widehat{\sigma}^2) db$$

$$\leq \exp\left\{ -e^{-\widetilde{M}+B} \widehat{\Lambda}(\tau - \epsilon) \right\} < \frac{\delta}{2}$$

and

$$n^{-1}\{l_n(\widehat{\boldsymbol{\theta}}, \widehat{\mathcal{A}}) - l_n(\boldsymbol{\theta}_0, \widetilde{\mathcal{A}})\}$$

$$< O(1) + n^{-1} \sum_{i=1}^{n} I\{\Delta_i G_i(1 - W_i) = 1, Q^*_{i,N_i+1} < \tau - \epsilon\} \log \delta$$

$$\to O(1) + P(\Delta_i G_i(1 - W_i) = 1, Q^*_{i,N_i+1} < \tau - \epsilon) \log \delta.$$

for infinitely many $n$. Since $\delta$ is arbitrary, $\liminf_n n^{-1}\{l_n(\widehat{\boldsymbol{\theta}}, \widehat{\mathcal{A}}) - l_n(\boldsymbol{\theta}_0, \widetilde{\mathcal{A}})\}$ will be negative, which contradicts the fact that it is always non-negative. Therefore, we conclude that $\limsup_n \widehat{V}(\tau) < \infty$ and $\limsup_n \widehat{\Lambda}(\tau - \epsilon) < \infty$ for any $\epsilon > 0$.

**Lemma S2.** *If $\ell_{\mathcal{A}}(\boldsymbol{\theta}_0, \mathcal{A}_0)(h) = 0$ with probability 1, then $h = 0$ almost surely in $[0, \tau] \times \mathcal{Q}$.*

**Proof** With probability 1,

$$\int_b J_1(\mathbf{O}_i, b; \theta_0, \mathcal{A}_0)\ell_{\mathcal{A}}(\boldsymbol{\theta}_0, \mathcal{A}_0)(h)\psi(b; \sigma_0^2)db = 0.$$

We set $\Delta_i = 0$. Using the same technique of proving identifiability in Lemma 1, we can obtain

$$\int_b \exp\left\{-\int_0^{Y_i} e^{\boldsymbol{\gamma}_0^\top \mathbf{X}_i(s)+b} dV_0(s)\right\}$$
$$\times \exp\left\{-\int_0^{Q_{il}} e^{\boldsymbol{\beta}_0^\top \mathbf{X}_i(s)+b} d\Lambda_0(s)\right\} \ell_{\mathcal{A}}^*(\boldsymbol{\theta}_0, \mathcal{A}_0)(h)\psi(b; \sigma_0^2)db = 0,$$

$$\int_b \exp\left\{-\int_0^{Y_i} e^{\boldsymbol{\gamma}_0^\top \mathbf{X}_i(s)+b} dV_0(s)\right\}$$
$$\times \left[1 - \exp\left\{-\int_0^{Q_{il}} e^{\boldsymbol{\beta}_0^\top \mathbf{X}_i(s)+b} d\Lambda_0(s)\right\}\right] \ell_{\mathcal{A}}^*(\boldsymbol{\theta}_0, \mathcal{A}_0)(h)\psi(b; \sigma_0^2)db = 0.$$

We add these two equations,

$$\int_b \exp\left\{-\int_0^{Y_i} e^{\boldsymbol{\gamma}_0^\top \mathbf{X}_i(s)+b} dV_0(s)\right\} \ell_{\mathcal{A}}^*(\boldsymbol{\theta}_0, \mathcal{A}_0)(h)\psi(b; \sigma_0^2)db = 0,$$

where

$$\ell_{\mathcal{A}}^*(\boldsymbol{\theta}_0, \mathcal{A}_0)(h) = \int_0^{Y_i} e^{\boldsymbol{\gamma}_0^\top \mathbf{X}_i(t)+b} h_1(t)dV_0(t) + \int_0^{Q_{il}} e^{\boldsymbol{\beta}_0^\top X_i(t)+b} h_2(t)d\Lambda_0(t).$$

Since $Y_i$ is arbitrary, we differentiate this equation with respect to $Y_i$, and we will have $h_1(t) = 0$. We can also differentiate this equation with respect to $Q_{il}$ on $\mathcal{Q}$, and we will have $h_2(t) = 0$.

**Lemma S3.** *If $\dot{\ell}(\boldsymbol{\theta}_0, \mathcal{A}_0) = 0$ with probability 1, then $\mathbf{v} = 0$.*

**Proof** Let $\mathbf{v} = (\mathbf{v}_1^\top, \mathbf{v}_2^\top, v_3)^\top$ be a vector correspond to $\boldsymbol{\beta}$, $\boldsymbol{\gamma}$ and $\sigma^2$. The score function

$$\dot{\ell}(\boldsymbol{\theta}_0, \mathcal{A}_0) = \Delta_i\{\mathbf{v}_1^\top \mathbf{X}_i(Y_i) - \mathbf{v}^\top \mathbf{h}_1^*(Y_i)\} - \int_0^\tau H(t, \mathbf{O}_i; \theta_0, \mathcal{A}_0)\{\mathbf{v}_1^\top \mathbf{X}_i(t) - \mathbf{v}^\top \mathbf{h}_1^*(t)\}dV_0(t)$$
$$+ \int_0^\tau H_2(t, \mathbf{O}_i; \theta_0, \mathcal{A}_0)\{\mathbf{v}_2^\top \mathbf{X}_i(t) - \mathbf{v}^\top \mathbf{h}_2^*(t)\}d\Lambda_0(t) + v_3 \frac{\int_b J_1(\mathbf{O}_i, b; \theta_0, \mathcal{A}_0)\dot{\psi}(b; \sigma_0^2)db}{\int_b J_1(\mathbf{O}_i, b; \theta_0, \mathcal{A}_0)\psi(b; \sigma_0^2)db}$$

is zero with probability 1, i.e.,

$$\int_b J_1(\mathbf{O}_i, b; \theta_0, \mathcal{A}_0)\left[\Delta_i\{\mathbf{v}_1^\top \mathbf{X}_i(Y_i) - \mathbf{v}^\top \mathbf{h}_1^*(Y_i)\} - \int_0^{Y_i} e^{\boldsymbol{\gamma}_0^\top \mathbf{X}_i(t)+b}\{\mathbf{v}_1^\top \mathbf{X}_i(t) - \mathbf{v}^\top \mathbf{h}_1^*(t)\}dV_0(t)\right]\psi(b; \sigma_0^2)db$$
$$+ \int_b \int_0^\tau \{J_3(t, \mathbf{O}_i, b; \theta_0, \mathcal{A}_0) - J_4(t, \mathbf{O}_i, b; \theta_0, \mathcal{A}_0)\}\{\mathbf{v}_2^\top \mathbf{X}_i(t) - \mathbf{v}^\top \mathbf{h}_2^*(t)\}d\Lambda_0(t)\psi(b; \sigma_0^2)db$$
$$+ v_3 \int_b J_1(\mathbf{O}_i, b; \theta_0, \mathcal{A}_0)\left\{-\frac{1}{2\sigma_0^2} + \frac{b^2}{2\sigma_0^4}\right\}\psi(b; \sigma_0^2)db = 0.$$

This equation can be written as

$$\int_b J_1(\mathbf{O}_i, b; \boldsymbol{\theta}_0, \mathcal{A}_0)\dot{\ell}^*(\boldsymbol{\theta}_0, \mathcal{A}_0)\psi(b; \sigma_0^2)db = 0.$$

We set $\Delta_i = 0$. Using the same technique of proving identifiability in Lemma 1, we can obtain

$$\int_b \exp\left\{-\int_0^{Y_i} e^{\gamma_0^\top \mathbf{X}_i(s)+b}dV_0(s)\right\} \exp\left\{-\int_0^{Q_{il}} e^{\boldsymbol{\beta}_0^\top \mathbf{X}_i(s)+b}d\Lambda_0(s)\right\}\dot{\ell}^*(\boldsymbol{\theta}_0, \mathcal{A}_0)\psi(b; \sigma_0^2)db = 0,$$

$$\int_b \exp\left\{-\int_0^{Y_i} e^{\gamma_0^\top \mathbf{X}_i(s)+b}dV_0(s)\right\}\left[1 - \exp\left\{-\int_0^{Q_{il}} e^{\boldsymbol{\beta}_0^\top \mathbf{X}_i(s)+b}d\Lambda_0(s)\right\}\right]\dot{\ell}^*(\boldsymbol{\theta}_0, \mathcal{A}_0)\psi(b; \sigma_0^2)db = 0.$$

We add these two equations,

$$\int_b \exp\left\{-\int_0^{Y_i} e^{\gamma_0^\top \mathbf{X}_i(s)+b}dV_0(s)\right\}\dot{\ell}^*(\boldsymbol{\theta}_0, \mathcal{A}_0)\psi(b; \sigma_0^2)db = 0,$$

where

$$\dot{\ell}(\boldsymbol{\theta}_0, \mathcal{A}_0) = \int_0^{Y_i} e^{\gamma_0^\top \mathbf{X}_i(t)+b}\{\mathbf{v}_1^\top \mathbf{X}_i(t) - \mathbf{v}^\top \mathbf{h}_1^*(t)\}dV_0(t)$$

$$+ \int_0^{Q_{il}} e^{\boldsymbol{\beta}_0^\top X_i(t)+b}\{\mathbf{v}_2^\top \mathbf{X}_i(t) - \mathbf{v}^\top \mathbf{h}_2^*(t)\}d\Lambda_0(t)$$

$$+ v_3\left\{-\frac{1}{2\sigma_0^2} + \frac{b^2}{2\sigma_0^4}\right\}.$$

Since $Y_i$ is arbitrary, we differentiate this equation with respect to $Y_i$, and we will have $\mathbf{v}_1^\top \mathbf{X}_i(t) - \mathbf{v}^\top \mathbf{h}_1^*(t) = 0$. By Condition 2, $\mathbf{v}_1 = 0$. We can also differentiate this equation with respect to $Q_{il}$ on $\mathcal{Q}$, and we will have $\mathbf{v}_2^\top \mathbf{X}_i(t) - \mathbf{v}^\top \mathbf{h}_2^*(t) = 0$. By Condition 2, $\mathbf{v}_2 = 0$. Then we apply the inverse Laplacian transformation and obtain

$$v_3\left\{-\frac{1}{2\sigma_0^2} + \frac{b^2}{2\sigma_0^4}\right\} = 0$$

for any $b$, so $v_3 = 0$.

**Lemma S4.** *The second-order terms $r_1$ and $r_2$ are bounded by $O_p(\sqrt{n}\|\widehat{\boldsymbol{\theta}} - \boldsymbol{\theta}_0\|^2 + n^{-1/6})$.*

**Proof** The second-order terms $r_1$ and $r_2$ are bounded by

$$O_p(1)\sqrt{n}\left[\|\widehat{\boldsymbol{\theta}} - \boldsymbol{\theta}_0\|^2 + \{\widehat{V}(Y) - V_0(Y)\}^2 + \sum_{l=1}^{N+1}\{\widehat{\Lambda}(Q_l^*) - \Lambda_0(Q_l^*)\}^2\right].$$

Since $m(\boldsymbol{\theta}, \mathcal{A})$ is Donsker, by Lemma 3.4.10 of van der Vaart and Wellner (2023),

$$\mathbb{P}\{m(\widehat{\boldsymbol{\theta}}, \widehat{\mathcal{A}}) - m(\boldsymbol{\theta}_0, \widetilde{\mathcal{A}})\} \leq -H^2\{(\widehat{\boldsymbol{\theta}}, \widehat{\mathcal{A}}), (\boldsymbol{\theta}_0, \widetilde{\mathcal{A}})\},$$

where

$$H\{(\widehat{\boldsymbol{\theta}}, \widehat{\mathcal{A}}), (\boldsymbol{\theta}_0, \widetilde{\mathcal{A}})\} = \left(\int \left[\sqrt{L(\widehat{\boldsymbol{\theta}}, \widehat{\mathcal{A}})} - \sqrt{L(\boldsymbol{\theta}_0, \widetilde{\mathcal{A}})}\right]^2 d\mu\right)^{1/2}$$

is the Hellinger distance with respect to a dominating measure $\mu$. From Lemma 2 of Gao et al. (2019), the $\epsilon$-bracketing number of $\{L_l(\boldsymbol{\theta}, \mathcal{A}) : \boldsymbol{\theta} \in \Theta, \mathcal{A} \in \mathcal{A}_M \times \mathcal{D}_M\}$ is $O(\exp\{\epsilon^{-2V/(V+2)} + \epsilon^{-1}\}\epsilon^{-2})$, where

$$L_l(\boldsymbol{\theta}, \mathcal{A}) = \nu\{Y_i\}^{\Delta_i} \int_b \exp\{\boldsymbol{\gamma}^\top \mathbf{X}_i(Y_i) + b\}^{\Delta_i} \exp\left\{-\int_0^{Y_i} e^{\boldsymbol{\gamma}^\top \mathbf{X}_i(s)+b} dV(s)\right\}$$

$$P(\boldsymbol{\xi}_i \mid T_i \in I_{il}) \left[\exp\left\{-\int_0^{Q_{il}^*} e^{\boldsymbol{\beta}^\top \mathbf{X}_i(s)+b} d\Lambda(s)\right\}\right.$$

$$\left. - I(Q_{i,l+1}^* < \infty)\exp\left\{-\int_0^{Q_{i,l+1}^*} e^{\boldsymbol{\beta}^\top \mathbf{X}_i(s)+b} d\Lambda(s)\right\}\right]\psi(b;\sigma^2)db$$

and $V \geq 2$ is the VC-index of $\{\boldsymbol{\beta}^\top \mathbf{X}_i(\cdot), \boldsymbol{\gamma}^\top \mathbf{X}_i(\cdot) : \boldsymbol{\theta} \in \Theta\}$. By the bracketing entropy of additive functions, the $\epsilon$-bracketing numbers of $\{L(\boldsymbol{\theta}, \mathcal{A}) : \boldsymbol{\theta} \in \Theta, \mathcal{A} \in \mathcal{D}_M \times \mathcal{D}_M\}$ and $\mathcal{M}$ are also $N_{[]}(\epsilon, \mathcal{M}, L_2(\mathbb{P})) = O(\exp\{\epsilon^{-2V/(V+2)} + \epsilon^{-1}\}\epsilon^{-2})$. Therefore,

$$\phi(\delta) = \int_0^\delta \sqrt{1 + \log N_{[]}(\epsilon, \mathcal{M}, L_2(\mathbb{P}))} d\epsilon \leq O(\delta^{2/(V+2)}) \leq O(\delta^{1/2}).$$

By Theorem 3.4.1 of van der Vaart and Wellner (2023), there exists $r_n$ such that $r_n^2 \phi(1/r_n) = O(\sqrt{n})$ and $H\{(\widehat{\boldsymbol{\theta}}, \widehat{\mathcal{A}}), (\boldsymbol{\theta}_0, \widetilde{\mathcal{A}})\} = O_p(r_n^{-1})$. In particular, we can choose $r_n = n^{1/3}$, which implies $H^2\{(\widehat{\boldsymbol{\theta}}, \widehat{\mathcal{A}}), (\boldsymbol{\theta}_0, \widetilde{\mathcal{A}})\} = O_p(n^{-2/3})$. By the mean value theorem,

$$O_p(n^{-2/3}) = E[L(\widehat{\boldsymbol{\theta}}, \widehat{\mathcal{A}}) - L(\boldsymbol{\theta}_0, \widetilde{\mathcal{A}})].$$

By the mean value theorem again,

$$O_p(n^{-2/3}) = E\left[\Delta_i(\widehat{\nu}\{Y_i\} - \widetilde{\nu}\{Y_i\}) - \int_0^\tau H(t, \mathbf{O}_i; \boldsymbol{\theta}_0, \widetilde{\mathcal{A}})\{d\widehat{V}(t) - d\widetilde{V}(t)\}\right.$$

$$\left. + \int_0^\tau H_2(t, \mathbf{O}_i; \boldsymbol{\theta}_0, \mathcal{A}_0)\{d\widehat{\Lambda}(t) - d\widetilde{\Lambda}(t)\}\right] + O_p(1)\|\widehat{\boldsymbol{\theta}} - \boldsymbol{\theta}_0\|$$

We define a norm

$$\|\widehat{\mathcal{A}} - \widetilde{\mathcal{A}}\|_1 = \left(E\left[\{\widehat{V}(Y_i) - \widetilde{V}(Y_i)\}^2 + \sum_{l=1}^{N_i+1}\{\widehat{\Lambda}(Q_{il}^*) - \widetilde{\Lambda}(Q_{il}^*)\}^2\right]\right)^{1/2}$$

and a seminorm

$$\|\widehat{\mathcal{A}} - \widetilde{\mathcal{A}}\|_2 = \left(E\left[\Delta_i(\widehat{\nu}\{Y_i\} - \widetilde{\nu}\{Y_i\}) - \int_0^\tau H(t, \mathbf{O}_i; \boldsymbol{\theta}_0, \mathcal{A}_0)\{d\widehat{V}(t) - d\widetilde{V}(t)\}\right.\right.$$

$$\left.\left. + \int_0^\tau H_2(t, \mathbf{O}_i; \boldsymbol{\theta}_0, \mathcal{A}_0)\{d\widehat{\Lambda}(t) - d\widetilde{\Lambda}(t)\}\right]^2\right)^{1/2}.$$

By observing the expression of $H_2(t, \mathbf{O}_i; \boldsymbol{\theta}_0, \mathcal{A}_0)$, we can decompose it into the sum of functions according to the mixture,

$$H_2(t, \mathbf{O}_i; \boldsymbol{\theta}_0, \mathcal{A}_0) = \sum_{l=1}^{N_i+1} H_{2l}(t, \mathbf{O}_i; \boldsymbol{\theta}_0, \mathcal{A}_0).$$

There is an indicator function $I(Y_i \leq t)$ in $H(t, \mathbf{O}_i; \boldsymbol{\theta}_0, \mathcal{A}_0)$ and an indicator function $I(Q_{il}^* \leq t)$ in $H_{2l}(t, \mathbf{O}_i; \boldsymbol{\theta}_o, \mathcal{A}_0)$. By the Cauchy–Schwarz inequality,

$$\|\widehat{\mathcal{A}} - \widetilde{\mathcal{A}}\|_2^2 = E\left[\int_0^\tau \{\Delta_i I(Y_i = t) - H(t, \mathbf{O}_i; \boldsymbol{\theta}_0, \mathcal{A}_0)\}\{d\widehat{V}(t) - d\widetilde{V}(t)\}\right.$$

$$+ \sum_{l=1}^{N_i+1} \int_0^\tau H_{2l}(t, \mathbf{O}_i; \boldsymbol{\theta}_0, \mathcal{A}_0)\{d\widehat{\Lambda}(t) - d\widetilde{\Lambda}(t)\}\Big]^2$$

$$\leq E\left[\int_0^\tau \left\{\Delta_i I(Y_i = t) - H(t, \mathbf{O}_i; \boldsymbol{\theta}_0, \mathcal{A}_0) + \sum_{l=1}^{N_i+1} H_{2l}(t, \mathbf{O}_i; \boldsymbol{\theta}_0, \mathcal{A}_0)\right\}^2 dt\right]$$

$$\times E\left[\{\widehat{V}(Y_i) - \widetilde{V}(Y_i)\}^2 + \sum_{l=1}^{N_i+1}\{\widehat{\Lambda}(Q^*_{il}) - \widetilde{\Lambda}(Q^*_{il})\}^2\right]$$

$$\leq c_1 \|\widehat{\mathcal{A}} - \widetilde{\mathcal{A}}\|_1^2,$$

where $c_1$ is a constant. In addition, if $\|f\|_2^2 = 0$, then

$$\Delta_i df_1(Y_i) - \int_0^\tau H(t, \mathbf{O}_i; \boldsymbol{\theta}_0, \mathcal{A}_0) df_1(t) + \int_0^\tau H_2(t, \mathbf{O}_i; \boldsymbol{\theta}_0, \mathcal{A}_0) df_2(t) = 0$$

with probability 1, i.e.,

$$\int_b J_1(\mathbf{O}_i, b; \boldsymbol{\theta}_0, \mathcal{A}_0)\left\{\Delta_i df_1(Y_i) - \int_0^{Y_i} e^{\gamma_0^\top \mathbf{X}_i(s)+b} df_1(t)\right\}\psi(b; \sigma_0^2)db$$
$$+ \int_b \int_0^\tau \{J_3(t, \mathbf{O}_i, b; \boldsymbol{\theta}_0, \mathcal{A}_0) - J_4(t, \mathbf{O}_i, b; \boldsymbol{\theta}_0, \mathcal{A}_0)\}df_2(t)\psi(b; \sigma_0^2)db = 0$$

with probability 1. To see this, we rewrite the equation above as

$$\int_b \nu(Y_i)^{\Delta_i} J_1(\mathbf{O}_i, b; \boldsymbol{\theta}_0, \mathcal{A}_0)\ell_\mathcal{A}(\boldsymbol{\theta}_0, \mathcal{A}_0)(f)\psi(b; \sigma_0^2)db = 0.$$

By Lemma S2, $f_1(t) = f_2(t) = 0$. This implies that $\|\cdot\|_2$ is a norm. By the inverse mapping theorem in Banach space, there exists a constant $c_2$ such that $\|\widehat{\mathcal{A}} - \widetilde{\mathcal{A}}\|_1^2 = \|\widehat{\mathcal{A}} - \mathcal{A}_0\|_1^2 \leq c_2 \|\widehat{\mathcal{A}} - \widetilde{\mathcal{A}}\|_2^2$. So the second-order term is bounded by

$$\sqrt{n} O_p(\|\widehat{\boldsymbol{\theta}} - \boldsymbol{\theta}_0\|^2 + \|\widehat{\mathcal{A}} - \widetilde{\mathcal{A}}\|_2^2) = \sqrt{n} O_p(\|\widehat{\boldsymbol{\theta}} - \boldsymbol{\theta}_0\|^2 + n^{-2/3})$$

# E  Additional Simulation Results

To assess the influence of sample size, Table S1 shows the summary of simulation results when $n = 1000$ under the convergence criterion tol $< 5 \times 10^{-5}$. Compared to the results with $n = 500$, the standard deviation is smaller. Estimation of the variance of the random effect $\sigma^2$ is more accurate when the sample size is large and when $p$ and $q$ are large. The standard deviation (SD) and standard error (SE) are in closer agreement.

To assess the influence of the convergence criterion, we set a stricter criterion that the maximum absolute difference of estimates in two adjacent iterations (tolerance) should be tol $< 1 \times 10^{-4}$. Table S2 shows the summary of estimation results. Compared with the results in the main text, the bias is generally slightly smaller, but the difference is not significant. However, the number of iterations to achieve convergence and the computation time increase significantly. We set the maximum number of iterations to 1000. In Setting 2, the median number of iterations to achieve convergence is 543.5, about 4.3 times that under the original convergence criterion. The computation time is multiplied accordingly.

To assess the influence of initial values in the algorithm, we randomly choose the initial values from a distribution $\beta_1 \sim N(0.5, 1)$, $\beta_2 \sim N(0.5, 1)$, $\gamma_1 \sim N(0.5, 1)$, $\gamma_2 \sim N(-0.5, 1)$, and $\sigma^2 \sim Gamma(2, 4)$

Table S1: Summary of the simulation results for $n = 1000$ under the convergence criterion $\text{tol} < 5 \times 10^{-4}$

| | Proposed method | | | | First-Diag | | | | Last-Diag | | | | NoRandEff | | | |
|---|---|---|---|---|---|---|---|---|---|---|---|---|---|---|---|---|
| | Bias | SD | SE | CP | Bias | SD | SE | CP | Bias | SD | SE | CP | Bias | SD | SE | CP |
| | Setting 1: $p = 0.9, q = 0.6, r = 1$ | | | | | | | | | | | | | | | |
| $\beta_1$ | .006 | .110 | .105 | .943 | -.370 | .070 | .069 | .000 | -.238 | .081 | .078 | .155 | -.032 | .100 | .092 | .917 |
| $\beta_2$ | .003 | .194 | .198 | .952 | -.231 | .126 | .127 | .549 | -.177 | .143 | .146 | .762 | -.043 | .178 | .146 | .883 |
| $\gamma_1$ | -.002 | .114 | .110 | .936 | -.019 | .110 | .109 | .939 | -.025 | .113 | .108 | .926 | -.024 | .109 | .099 | .915 |
| $\gamma_2$ | .006 | .193 | .188 | .944 | .024 | .185 | .186 | .951 | .029 | .190 | .185 | .937 | .027 | .184 | .144 | .867 |
| $\sigma^2$ | -.005 | .090 | .095 | .964 | -.203 | .014 | .064 | .000 | -.137 | .026 | .081 | .771 | – | – | – | – |
| | Setting 2: $p = 0.9, q = 0.6, r = 0.5$ | | | | | | | | | | | | | | | |
| $\beta_1$ | .005 | .117 | .113 | .941 | -.354 | .074 | .069 | .001 | -.249 | .088 | .085 | .182 | -.036 | .107 | .098 | .910 |
| $\beta_2$ | .004 | .210 | .211 | .951 | -.311 | .128 | .128 | .313 | -.211 | .157 | .158 | .724 | -.032 | .194 | .152 | .866 |
| $\gamma_1$ | -.002 | .114 | .110 | .934 | -.015 | .111 | .108 | .935 | -.022 | .114 | .109 | .926 | -.024 | .109 | .099 | .914 |
| $\gamma_2$ | .006 | .193 | .188 | .942 | .020 | .187 | .184 | .947 | .025 | .191 | .186 | .938 | .027 | .184 | .147 | .873 |
| $\sigma^2$ | -.004 | .095 | .106 | .965 | -.153 | .039 | .066 | .277 | -.099 | .051 | .095 | .991 | – | – | – | – |
| | Setting 3: $p = 0.9, q = 0.6, r = 0$ | | | | | | | | | | | | | | | |
| $\beta_1$ | .005 | .127 | .123 | .941 | -.299 | .089 | .085 | .067 | -.167 | .120 | .112 | .649 | -.042 | .117 | .106 | .911 |
| $\beta_2$ | .007 | .229 | .228 | .946 | -.394 | .153 | .155 | .278 | -.271 | .206 | .206 | .728 | -.013 | .213 | .160 | .861 |
| $\gamma_1$ | -.002 | .114 | .110 | .935 | .019 | .121 | .116 | .931 | .033 | .133 | .122 | .919 | -.024 | .109 | .099 | .916 |
| $\gamma_2$ | .005 | .194 | .188 | .941 | -.015 | .202 | .195 | .938 | -.030 | .217 | .208 | .942 | .027 | .184 | .149 | .881 |
| $\sigma^2$ | .004 | .125 | .123 | .940 | .250 | .195 | .094 | .220 | .502 | .376 | .148 | .071 | – | – | – | – |
| | Setting 4: $p = 0.8, q = 0.5, r = 1$ | | | | | | | | | | | | | | | |
| $\beta_1$ | .013 | .137 | .129 | .933 | -.412 | .069 | .068 | .000 | -.304 | .079 | .078 | .027 | -.026 | .124 | .109 | .911 |
| $\beta_2$ | .009 | .241 | .245 | .954 | -.266 | .125 | .126 | .411 | -.252 | .139 | .145 | .580 | -.042 | .219 | .165 | .850 |
| $\gamma_1$ | -.002 | .115 | .110 | .935 | -.020 | .110 | .109 | .939 | -.025 | .113 | .108 | .929 | -.024 | .109 | .100 | .915 |
| $\gamma_2$ | .005 | .193 | .188 | .941 | .025 | .185 | .187 | .951 | .029 | .190 | .185 | .939 | .027 | .184 | .153 | .890 |
| $\sigma^2$ | .001 | .104 | .100 | .947 | -.209 | .013 | .066 | .000 | -.138 | .025 | .081 | .800 | – | – | – | – |
| | Setting 5: $p = 0.8, q = 0.5, r = 0.5$ | | | | | | | | | | | | | | | |
| $\beta_1$ | .011 | .159 | .150 | .942 | -.400 | .072 | .069 | .000 | -.326 | .086 | .085 | .032 | -.028 | .144 | .127 | .916 |
| $\beta_2$ | .012 | .282 | .284 | .949 | -.360 | .131 | .126 | .195 | -.297 | .153 | .157 | .514 | -.034 | .258 | .182 | .828 |
| $\gamma_1$ | -.002 | .115 | .110 | .936 | -.016 | .111 | .108 | .937 | -.022 | .114 | .109 | .928 | -.024 | .109 | .101 | .920 |
| $\gamma_2$ | .005 | .194 | .188 | .940 | .020 | .187 | .184 | .944 | .026 | .191 | .186 | .941 | .027 | .184 | .158 | .899 |
| $\sigma^2$ | .000 | .116 | .118 | .955 | -.162 | .036 | .068 | .222 | -.100 | .054 | .096 | .989 | – | – | – | – |
| | Setting 6: $p = 0.8, q = 0.5, r = 0$ | | | | | | | | | | | | | | | |
| $\beta_1$ | .011 | .198 | .186 | .941 | -.355 | .088 | .086 | .019 | -.267 | .120 | .113 | .324 | -.038 | .182 | .157 | .909 |
| $\beta_2$ | .025 | .341 | .348 | .952 | -.471 | .157 | .158 | .144 | -.394 | .208 | .209 | .512 | .004 | .318 | .208 | .815 |
| $\gamma_1$ | -.001 | .117 | .110 | .933 | .023 | .122 | .117 | .931 | .039 | .134 | .123 | .914 | -.024 | .109 | .102 | .921 |
| $\gamma_2$ | .004 | .196 | .189 | .937 | -.019 | .204 | .197 | .940 | -.036 | .221 | .211 | .945 | .027 | .184 | .163 | .912 |
| $\sigma^2$ | .010 | .198 | .154 | .916 | .295 | .203 | .105 | .184 | .566 | .403 | .153 | .019 | – | – | – | – |

instead of constants. Table S3 shows the summary of simulation results under such initial values. The standard deviation, standard error, and coverage percentages are similar to those in the main text, where initial values are chosen as fixed constants. The results indicate that estimation is insensitive to initial values.

Table S2: Summary of the simulation results for $n = 500$ under the convergence criterion tol $< 1 \times 10^{-4}$

| | Proposed method | | | | First-Diag | | | | Last-Diag | | | | NoRandEff | | | |
|---|---|---|---|---|---|---|---|---|---|---|---|---|---|---|---|---|
| | Bias | SD | SE | CP | Bias | SD | SE | CP | Bias | SD | SE | CP | Bias | SD | SE | CP |
| | Setting 1: $p = 0.9, q = 0.6, r = 1$ | | | | | | | | | | | | | | | |
| $\beta_1$ | .013 | .153 | .153 | .951 | -.377 | .094 | .097 | .027 | -.245 | .108 | .110 | .381 | -.029 | .138 | .131 | .935 |
| $\beta_2$ | .025 | .298 | .283 | .935 | -.245 | .179 | .178 | .698 | -.180 | .200 | .202 | .852 | -.026 | .269 | .206 | .860 |
| $\gamma_1$ | .002 | .161 | .161 | .953 | -.020 | .153 | .156 | .957 | -.028 | .156 | .154 | .946 | -.021 | .152 | .140 | .929 |
| $\gamma_2$ | .002 | .264 | .269 | .958 | .025 | .252 | .263 | .961 | .034 | .259 | .261 | .954 | .026 | .251 | .204 | .879 |
| $\sigma^2$ | .014 | .160 | .170 | .980 | -.240 | .004 | .100 | .001 | -.196 | .047 | .144 | .975 | – | – | – | – |
| | Setting 2: $p = 0.9, q = 0.6, r = 0.5$ | | | | | | | | | | | | | | | |
| $\beta_1$ | .016 | .163 | .165 | .950 | -.362 | .096 | .097 | .047 | -.257 | .116 | .122 | .428 | -.032 | .147 | .140 | .930 |
| $\beta_2$ | .025 | .318 | .304 | .946 | -.325 | .182 | .178 | .539 | -.212 | .218 | .219 | .834 | -.018 | .289 | .216 | .853 |
| $\gamma_1$ | .004 | .161 | .162 | .953 | -.016 | .154 | .154 | .950 | -.024 | .158 | .155 | .943 | -.021 | .152 | .141 | .931 |
| $\gamma_2$ | .001 | .266 | .270 | .956 | .022 | .254 | .259 | .958 | .030 | .261 | .264 | .958 | .026 | .251 | .208 | .891 |
| $\sigma^2$ | .033 | .188 | .201 | .984 | -.204 | .048 | .109 | .372 | -.158 | .090 | .188 | .999 | – | – | – | – |
| | Setting 3: $p = 0.9, q = 0.6, r = 0$ | | | | | | | | | | | | | | | |
| $\beta_1$ | .024 | .180 | .182 | .952 | -.304 | .122 | .121 | .301 | -.195 | .155 | .181 | .777 | -.037 | .160 | .151 | .931 |
| $\beta_2$ | .025 | .346 | .330 | .945 | -.416 | .225 | .221 | .526 | -.267 | .277 | .285 | .859 | -.003 | .316 | .226 | .846 |
| $\gamma_1$ | .009 | .163 | .164 | .955 | .020 | .167 | .165 | .942 | .021 | .175 | .181 | .965 | -.021 | .152 | .141 | .932 |
| $\gamma_2$ | -.004 | .268 | .273 | .956 | -.016 | .275 | .277 | .956 | -.016 | .286 | .297 | .965 | .026 | .251 | .211 | .895 |
| $\sigma^2$ | .085 | .241 | .265 | .987 | .222 | .216 | .191 | .844 | .324 | .329 | .675 | .988 | – | – | – | – |
| | Setting 4: $p = 0.8, q = 0.5, r = 1$ | | | | | | | | | | | | | | | |
| $\beta_1$ | .033 | .201 | .189 | .943 | -.415 | .092 | .096 | .011 | -.312 | .107 | .109 | .163 | -.019 | .181 | .157 | .915 |
| $\beta_2$ | .044 | .373 | .355 | .940 | -.272 | .178 | .177 | .644 | -.256 | .205 | .201 | .733 | -.012 | .337 | .236 | .837 |
| $\gamma_1$ | .004 | .162 | .160 | .950 | -.025 | .153 | .157 | .956 | -.032 | .158 | .154 | .944 | -.026 | .153 | .143 | .935 |
| $\gamma_2$ | .000 | .266 | .270 | .953 | .032 | .257 | .265 | .954 | .044 | .267 | .262 | .939 | .032 | .257 | .218 | .900 |
| $\sigma^2$ | .034 | .198 | .173 | .946 | -.241 | .003 | .099 | .000 | -.196 | .043 | .142 | .970 | – | – | – | – |
| | Setting 5: $p = 0.8, q = 0.5, r = 0.5$ | | | | | | | | | | | | | | | |
| $\beta_1$ | .040 | .233 | .222 | .943 | -.399 | .095 | .096 | .025 | -.331 | .122 | .120 | .187 | -.018 | .212 | .181 | .913 |
| $\beta_2$ | .057 | .440 | .414 | .937 | -.376 | .184 | .176 | .423 | -.299 | .227 | .219 | .722 | -.020 | .383 | .259 | .822 |
| $\gamma_1$ | .006 | .162 | .161 | .948 | -.008 | .156 | .154 | .950 | -.016 | .164 | .156 | .946 | -.011 | .154 | .144 | .946 |
| $\gamma_2$ | -.002 | .268 | .271 | .954 | .031 | .251 | .261 | .971 | .038 | .260 | .266 | .950 | .034 | .249 | .225 | .938 |
| $\sigma^2$ | .053 | .233 | .209 | .952 | -.210 | .035 | .107 | .303 | -.151 | .090 | .187 | 1.00 | – | – | – | – |
| | Setting 6: $p = 0.8, q = 0.5, r = 0$ | | | | | | | | | | | | | | | |
| $\beta_1$ | .065 | .300 | .283 | .951 | -.346 | .119 | .122 | .200 | -.284 | .155 | .175 | .575 | -.016 | .241 | .223 | .920 |
| $\beta_2$ | .062 | .554 | .513 | .934 | -.486 | .229 | .224 | .405 | -.382 | .284 | .291 | .740 | .003 | .481 | .295 | .770 |
| $\gamma_1$ | .016 | .168 | .166 | .951 | .031 | .169 | .166 | .950 | .038 | .180 | .186 | .970 | -.011 | .151 | .145 | .945 |
| $\gamma_2$ | -.010 | .272 | .278 | .961 | -.004 | .277 | .279 | .955 | -.009 | .293 | .302 | .965 | .036 | .255 | .233 | .925 |
| $\sigma^2$ | .166 | .378 | .374 | .961 | .225 | .197 | .193 | .850 | .403 | .380 | .716 | .985 | – | – | – | – |

## F  Additional Results for Data Application

Figure S1 presents the estimated cumulative incidence function of Alzheimer's disease and survival probability by the proposed method and competing methods (First-Diag, Last-Diag, NoRandEff), categorized by sex.

13

Table S3: Summary of the simulation results for $n = 500$ using random initial values under the convergence criterion tol $< 5 \times 10^{-4}$

|  | Proposed method | | | | First-Diag | | | | Last-Diag | | | | NoRandEff | | | |
|---|---|---|---|---|---|---|---|---|---|---|---|---|---|---|---|---|
|  | Bias | SD | SE | CP | Bias | SD | SE | CP | Bias | SD | SE | CP | Bias | SD | SE | CP |
| Setting 1: $p = 0.9, q = 0.6, r = 1$ | | | | | | | | | | | | | | | | |
| $\beta_1$ | .015 | .153 | .152 | .950 | -.372 | .097 | .098 | .032 | -.237 | .111 | .112 | .415 | -.030 | .138 | .131 | .935 |
| $\beta_2$ | .027 | .299 | .285 | .938 | -.238 | .184 | .180 | .728 | -.171 | .205 | .208 | .863 | -.027 | .268 | .207 | .866 |
| $\gamma_1$ | .004 | .161 | .158 | .948 | -.017 | .154 | .154 | .951 | -.022 | .158 | .155 | .947 | -.013 | .282 | .140 | .929 |
| $\gamma_2$ | .001 | .265 | .270 | .956 | .023 | .253 | .265 | .961 | .028 | .262 | .265 | .959 | .026 | .251 | .205 | .882 |
| $\sigma^2$ | .038 | .146 | .141 | .953 | -.206 | .013 | .102 | .417 | -.132 | .044 | .116 | .993 | – | – | – | – |
| Setting 2: $p = 0.9, q = 0.6, r = 0.5$ | | | | | | | | | | | | | | | | |
| $\beta_1$ | .017 | .162 | .164 | .947 | -.355 | .100 | .099 | .056 | -.248 | .119 | .122 | .461 | -.033 | .146 | .140 | .931 |
| $\beta_2$ | .026 | .317 | .305 | .947 | -.319 | .189 | .182 | .576 | -.207 | .224 | .225 | .841 | -.020 | .288 | .216 | .853 |
| $\gamma_1$ | .005 | .161 | .158 | .949 | -.012 | .155 | .153 | .950 | -.019 | .159 | .156 | .947 | .024 | .700 | .140 | .924 |
| $\gamma_2$ | -.001 | .266 | .270 | .956 | .018 | .256 | .261 | .958 | .024 | .264 | .267 | .958 | .027 | .255 | .208 | .886 |
| $\sigma^2$ | .052 | .170 | .160 | .951 | -.155 | .048 | .109 | .864 | -.094 | .080 | .142 | .992 | – | – | – | – |
| Setting 3: $p = 0.9, q = 0.6, r = 0$ | | | | | | | | | | | | | | | | |
| $\beta_1$ | .025 | .177 | .179 | .951 | -.304 | .121 | .120 | .293 | -.195 | .157 | .153 | .706 | -.039 | .159 | .151 | .932 |
| $\beta_2$ | .025 | .344 | .331 | .945 | -.415 | .226 | .221 | .524 | -.269 | .275 | .278 | .843 | -.003 | .314 | .227 | .852 |
| $\gamma_1$ | .009 | .163 | .160 | .947 | .020 | .167 | .163 | .941 | .020 | .177 | .169 | .946 | -.013 | .282 | .141 | .932 |
| $\gamma_2$ | -.005 | .268 | .272 | .954 | -.015 | .274 | .277 | .958 | -.015 | .288 | .289 | .951 | .026 | .251 | .211 | .895 |
| $\sigma^2$ | .098 | .213 | .188 | .932 | .221 | .204 | .149 | .669 | .336 | .367 | .209 | .622 | – | – | – | – |
| Setting 4: $p = 0.8, q = 0.5, r = 1$ | | | | | | | | | | | | | | | | |
| $\beta_1$ | .033 | .201 | .189 | .944 | -.415 | .094 | .097 | .010 | -.304 | .113 | .111 | .218 | -.019 | .177 | .157 | .917 |
| $\beta_2$ | .044 | .371 | .358 | .945 | -.272 | .182 | .179 | .657 | -.245 | .206 | .207 | .772 | -.023 | .326 | .236 | .849 |
| $\gamma_1$ | .006 | .162 | .159 | .950 | -.017 | .154 | .155 | .953 | -.022 | .158 | .155 | .948 | -.013 | .282 | .142 | .932 |
| $\gamma_2$ | -.002 | .267 | .271 | .956 | .023 | .253 | .266 | .959 | .029 | .263 | .265 | .956 | .026 | .251 | .217 | .911 |
| $\sigma^2$ | .061 | .178 | .151 | .925 | -.210 | .012 | .103 | .392 | -.132 | .044 | .117 | .992 | – | – | – | – |
| Setting 5: $p = 0.8, q = 0.5, r = 0.5$ | | | | | | | | | | | | | | | | |
| $\beta_1$ | .025 | .178 | .179 | .951 | -.304 | .121 | .120 | .294 | -.195 | .156 | .153 | .706 | -.039 | .159 | .151 | .932 |
| $\beta_2$ | .025 | .345 | .331 | .945 | -.415 | .226 | .221 | .524 | -.269 | .275 | .278 | .843 | -.003 | .314 | .227 | .852 |
| $\gamma_1$ | .009 | .163 | .160 | .947 | .020 | .167 | .163 | .941 | .020 | .177 | .169 | .946 | -.013 | .283 | .141 | .932 |
| $\gamma_2$ | -.004 | .268 | .272 | .954 | -.015 | .274 | .277 | .958 | -.015 | .288 | .289 | .951 | .026 | .252 | .211 | .895 |
| $\sigma^2$ | .097 | .213 | .188 | .932 | .221 | .204 | .149 | .668 | .335 | .367 | .209 | .623 | – | – | – | – |
| Setting 6: $p = 0.8, q = 0.5, r = 0$ | | | | | | | | | | | | | | | | |
| $\beta_1$ | .039 | .230 | .221 | .945 | -.401 | .097 | .098 | .021 | -.325 | .121 | .121 | .228 | -.016 | .204 | .182 | .917 |
| $\beta_2$ | .053 | .432 | .417 | .943 | -.364 | .190 | .181 | .460 | -.292 | .227 | .224 | .738 | -.014 | .384 | .259 | .823 |
| $\gamma_1$ | .007 | .162 | .159 | .947 | -.013 | .155 | .153 | .949 | -.019 | .160 | .156 | .947 | .024 | .700 | .143 | .929 |
| $\gamma_2$ | -.003 | .268 | .272 | .953 | .018 | .256 | .261 | .956 | .025 | .265 | .267 | .959 | .027 | .255 | .224 | .918 |
| $\sigma^2$ | .073 | .203 | .180 | .939 | -.162 | .041 | .110 | .852 | -.095 | .078 | .144 | .997 | – | – | – | – |

14

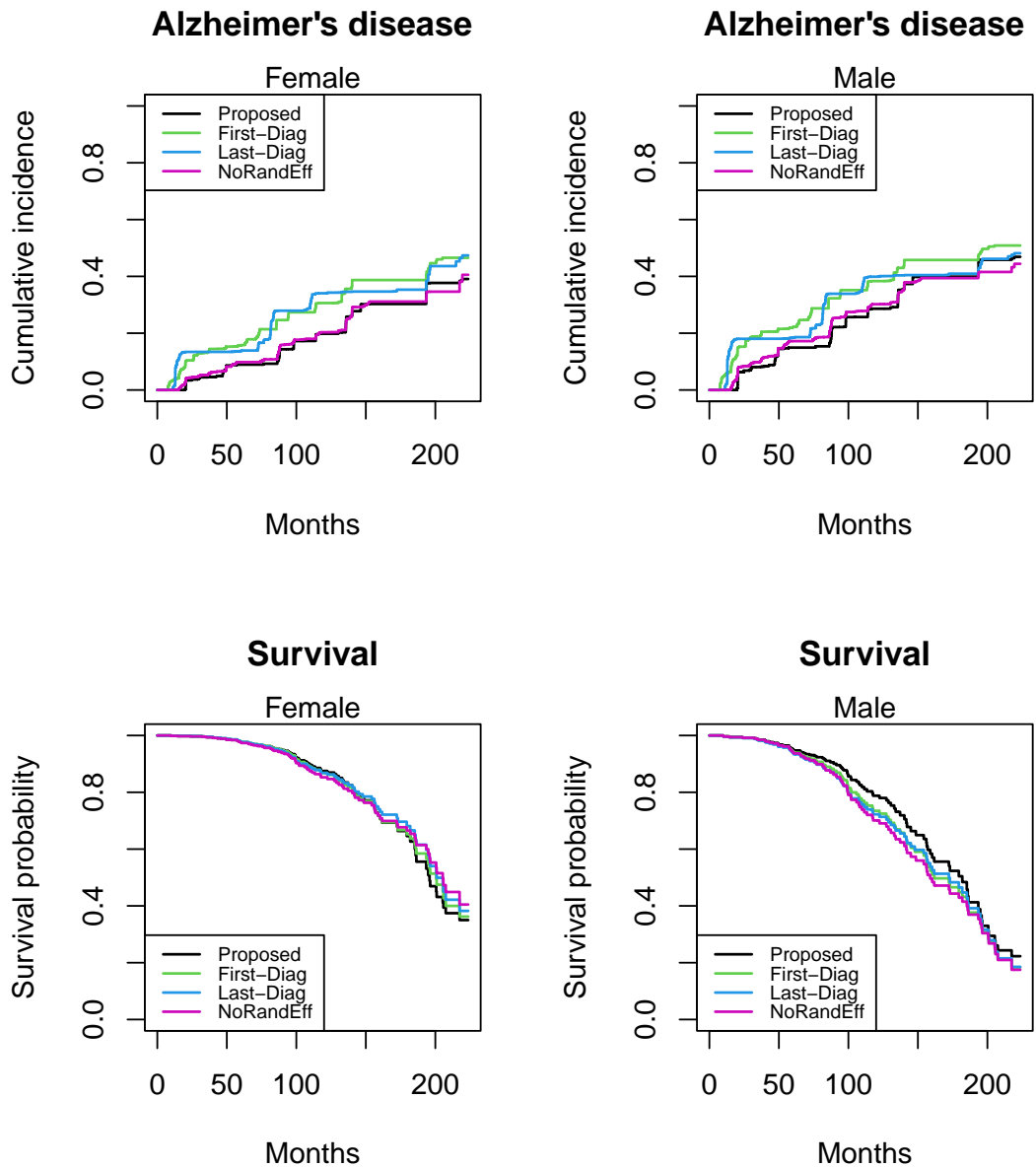

Figure S1: Estimated cumulative incidence function of Alzheimer's disease and survival probability by the proposed method and competing methods categorized by sex.

15



\end{document}